\documentclass[11pt,english,twoside]{article}
\usepackage[english]{babel}
\usepackage{amsmath,amsfonts}
\usepackage{mathtools}
\usepackage[cp850]{inputenc}
\usepackage[square]{natbib}%

\usepackage{tikz}
\usepackage{multirow}
\usepackage{dsfont}
\usepackage[affil-it]{authblk}
\usepackage[english]{babel}
\usepackage{blindtext}

\allowdisplaybreaks
\newcounter{theorem}[section]
\renewcommand{\thetheorem}{\thesection.\arabic{theorem}}
\newenvironment{lemma}[      1]{\refstepcounter{theorem} %
\bf \ Lemma#1. \thetheorem.       \it}{}
\newenvironment{assumption}[      1]{\refstepcounter{theorem} %
\bf \ Assumption#1. \thetheorem.       \it}{}
\newenvironment{theorem}[    1]{\refstepcounter{theorem} %
\bf \ Theorem#1. \thetheorem.     \it}{}

\newenvironment{corollary}[  1]{\refstepcounter{theorem} %
\bf \ Corollary#1. \thetheorem.   \it}{}

\newenvironment{example}[    1]{\refstepcounter{theorem} %
\bf \ Example#1. \thetheorem.     \rm}{\hfill $ \Box $}

\newenvironment{remark}[  1]{\refstepcounter{theorem} %
\bf \ Remark#1. \thetheorem.      \rm}{\hfill $ \Box $}

\newenvironment{proof}{ %
\it              Proof.          \rm}{\hfill $ \Box $}
\newenvironment{model}[    1]{\refstepcounter{theorem} %
\bf \ Model#1. \thetheorem.     \rm}{}

\newcounter{abc}[theorem]

\newcounter{one}[theorem]

\newcounter{rom}
\newenvironment{romlist}{\begin{list}{%
\rm (\roman{rom})  \hfill           }{\usecounter{rom} %
\topsep0mm \partopsep0mm \parsep0mm \itemsep0mm %
\leftmargin2em \labelwidth2em \labelsep0em}}{\end{list}}

\newenvironment{bulllist}{\begin{list}{%
$\bullet$               \hfill            }{%
\topsep0mm \partopsep0mm \parsep0mm \itemsep0mm %
\leftmargin2em \labelwidth2em \labelsep0em}}{\end{list}}

\newcommand{\Real}{{\mathbb{R}}}
\def\eqd{\buildrel {\rm d} \over =}
\def\P#1{{\mathbb{P}}\left[\{#1\right\}]}
\def\iid{\buildrel {\rm i.i.d.} \over \sim}
\def\E#1{{E}\left[#1\right]}
\def\Var#1{{\mathrm{Var}}\left[#1\right]}

\newcommand{\GGG}{{\bf G}}
\newcommand{\YYY}{{\bf Y}}

\newcommand{\ttt}{{\bf t}}
\newcommand{\vvv}{{\bf v}}
\newcommand{\xxx}{{\bf x}}
\newcommand{\yyy}{{\bf y}}
\def\Real{\mathbb{R}}
\newcommand{\I}{\mathbb{I}}

    \makeatletter
\def\@fnsymbol#1{\ensuremath{\ifcase#1\or \dagger\or \ddagger\or
   \mathsection\or \mathparagraph\or \|\or **\or \dagger\dagger
   \or \ddagger\ddagger \else\@ctrerr\fi}}
    \makeatother

\begin{document}

\title{\Large \bf A copula transformation in multivariate mixed discrete-continuous models}
\author[1]{Jae Youn Ahn\thanks{jaeyahn@ewha.ac.kr}}
\author[2,**]{Sebastian Fuchs\thanks{sebastian.fuchs@sbg.ac.at}}
\author[3,**]{Rosy Oh\thanks{rosy.oh5@gmail.com}}

\affil[1]{Ewha Womans University, Republic of Korea}
\affil[2]{Universit{\"a}t Salzburg, Austria}
\affil[3]{Ewha Womans University, Republic of Korea}
\affil[**]{Corresponding authors}

\date{}
\maketitle

\begin{abstract}
Copulas allow a flexible and simultaneous modeling of complicated dependence structures together with various marginal distributions. Especially if the density function can be represented as the product of the marginal density functions and the copula density function, this leads to both an intuitive interpretation of the conditional distribution and convenient estimation procedures.
However, this is no longer the case for copula models with mixed discrete and continuous marginal distributions,
because the corresponding density function cannot be decomposed so nicely.
In this paper, we introduce a copula transformation method that allows to represent the density function of a distribution with mixed discrete and continuous marginals
as the product of the marginal probability mass/density functions and the copula density function.
With the proposed method, conditional distributions can be described analytically and the computational complexity in the estimation procedure can be reduced depending on the type of copula used.
\end{abstract}


\section{Introduction}

Along with random effect methods, copula methods are a widely used tool to model multivariate distributions.
In case of both, the univariate marginal distribution functions and the copula associated with a $(d+1)$-dimensional distribution function $H$, are absolutely continuous, the density $h$ of $H$ fulfills
\begin{equation}\label{int.eq.1}
  h (x,y_1, \dots, y_d)
	= c \big( F(x), G_1(y_1), \dots, G_d(y_d) \big) \, f(x) \, \prod_{i=1}^{d} g_i(y_i),
\end{equation}
where $c$ denotes the copula density,
$F$ and $G_i$ denote the univariate marginal distribution functions and $f$ and $g_i$ denote its corresponding density functions, $i \in \{1,\dots,d\}$.
The copula representation (\ref{int.eq.1}) enables the dependence structure to be separated from the marginal distributions.
Such a complete separation not only provides a meaningful interpretation of the model
but also allows a convenient estimation \citep{Nelson, Joe3, durante2015principles}.
However, this convenience is lost when discrete distributions appear in the model.
Then, in contrast to (\ref{int.eq.1}),
the copula representation of the multivariate distribution no longer provides a complete separation of the dependence structure from the marginal distributions in the density function.
Thus, the interpretation of the dependence structure becomes difficult \citep{Genest6},
and the traditional statistical estimation procedures
cannot be directly applied \citep{Song2007}.
For instance,
when modeling $d$ discrete random variables, the evaluation of the likelihood function requires the calculation of $2^d$ terms\footnote{
When pair copula construction is adapted, the computational burden of evaluating $n$-dimensional discrete random variables only requires $2n(n-1)$ terms \citep{panagiotelis2012pair}.
}
which provoke computational difficulties in the estimation procedure and complicates an interpretation of the dependence structure \citep{Smith2012, Zilko2016}.
Copula methods for mixed - discrete and continuous - marginals, mixed copula models for short,
suffer from similar difficulties.

\bigskip
In this paper we investigate mixed copula models with a single discrete and several absolutely continuous variables and mainly focus on problems
related to interpretation difficulties of the (conditional) dependence structure and computational difficulties in estimation.

\bigskip
Often the dependence structure in the copula model is explained in terms of conditional distributions.
For instance,
a wide range of copula families including Archimedean copulas and some elliptical copula families are closed under the operation of conditioning \citep{Mesfioui2008,  Ding2016}.
For such copula families, the conditional distribution has both an analytical and an intuitive interpretation.
However, this convenience is lost in copula models with discrete marginals.
We note that the interpretability of conditional distributions when conditioning with respect to a discrete random variable is important in several research areas
including case-control studies in medicine \citep{He2012, de2011copula} and
frequency-severity models in insurance \citep{Czado, Kramer2013}.

\bigskip
A second problem that occurs in mixed copula models is the computational complexity in the calculation of density functions \citep{kadhem2019factor}
since statistical estimation procedures require the evaluation of the corresponding joint density function for multiple terms.
In the case of implicit copula models, for instance,
the calculation of the density function requires numerical integration
\citep{Nikoloulopoulos, kadhem2019factor}.
Such numerical difficulties may complicate the estimation procedure mainly due to the computational difficulties in calculating the likelihood functions and subsequent derivatives \citep{Song2007}.

\bigskip
In the present paper we aim at providing a handy representation of the mixed copula model so that both interpretation and calculation of the density function remain intuitive and simple.

To this end,
we start with a rather naive question:
Can we reformulate a mixed copula model with some discrete distribution function $F$ and some absolutely continuous distribution functions $G_1, \dots, G_n$, and find some closely related distribution whose density satisfies \eqref{int.eq.1}?
An answer to that question
requires a copula transformation that is presented in Section \ref{Sec.MM.}.
With the rather appealing form of the density in \eqref{int.eq.1},
we expect that the proposed distribution may provide a meaningful interpretation and excellence in computation in the estimation procedure.
A numerical analysis comparing the mixed copula model with the modified version is presented in Section \ref{Sec.NA.}.
As an application, we apply the proposed method to the collective risk model, CRM for short, (Section \ref{Sec.CRM.}) which plays a crucial role in insurance.
The CRM models the aggregate claim amount of a portfolio where the number of claims is random.
In particular, for the prediction of the fair premium, modeling the dependence structure in the CRM is important.
There are several ways to do this, whereby in this paper we focus on two CRMs that were presented in the recent actuarial literature:
two part frequency-severity model (see, e.g., \citet{Frees2, Peng, Garrido, Ahn2008}) and
copula-based CRM (see, e.g., \citet{cossette2019collective, oh2020copulaSAJ}).
These two methods were developed independently in different mathematical settings which makes the comparison of the two models difficult.
However, with the proposed copula transformation applied to the copula-based CRM,
we provide an example demonstrating the linkage between the two models.

\bigskip
Throughout the paper we write
$\I := [0,1]$,
$\mathbb{N}_0:=\mathbb{N} \cup \{0\}$
and let $d \geq 2$ be an integer which will be kept fixed.
Bold symbols are used for vectors, e.g., $\yyy = (y_1, \dots y_d) \in \mathbb{R}^d$, or vectors of functions, e.g., ${\bf f} = (f_1, \dots, f_d)$.
We denote by $\zeta^d$ the $d$-dimensional Lebesgue measure;
in case of $d = 1$ we simply write $\zeta$.


\section{Transformation of the mixed model}
\label{Sec.MM.}

In this section, we consider a multivariate mixed model with a single discrete and several absolutely continuous variables and present a modification of this model that allows both a meaningful interpretation and a convenient estimation.

\bigskip
First, fix some probability space $(\Omega, \mathcal{A}, \mathbb{P})$
and consider a random variable $N$ (on this probability space) with distribution function $F$
such that
$N$ follows a discrete distribution on $\mathbb{N}_0$.
Additionally, we consider a $d$-dimensional random vector $\YYY$ (on the same probability space) whose margins $Y_i$, $i \in \{1,\dots,d\}$,
follow an absolutely continuous distribution with $\zeta$--\,densities $g_i$
and distribution functions $G_i$, $i \in \{1,\dots,d\}$.
Then,
by Sklar's Theorem (see, e.g., \cite{dus2016, nel2006}), there exists some
$(d+1)$-dimensional copula $C$ such that the distribution function $H$ of $(N,\YYY)$
satisfies
\begin{equation}\label{Sklar}
  H (n,\yyy)	= C \big( F(n), \GGG(\yyy) \big)
\end{equation}
for every $(n,\yyy) \in \mathbb{N}_0 \times \mathbb{R}^d$.
Note that the copula $C$ fails to be unique, in general.
In the following,
we assume that $C$ is absolutely continuous, i.e.
there exists some $\zeta^{d+1}$-\,density $c$ of $C$.
Then, $H$ has $(\mu \otimes \zeta^{d})$-\,density
(where $\mu$ denotes the counting measure on the power set of $\mathbb{N}_0$)
$h$ satisfying
\begin{eqnarray}
  h (n,\yyy)
	& = & \left( \frac{\partial^d}{\partial \yyy} \, C \big( F(n), \GGG(\yyy) \big)
					 - \frac{\partial^d}{\partial \yyy} \, C \big( F(n-1), \GGG(\yyy) \big) \right)
					 \; \prod_{i=1}^{d} g_i(y_i) \label{s2.eq.2}
	\\
	& = & \left( \int_{(F(n-1),F(n)]}
					c \big( u, \GGG(\yyy) \big) \; \mathrm{d} \zeta(u) \right)
					 \; \prod_{i=1}^{d} g_i(y_i) \nonumber
\end{eqnarray}
for $(\mu \otimes \zeta^{d})$--\,almost
all $(n, \yyy) \in \mathbb{N}_0 \times \Real^d$.
While a model with a density function fulfilling \eqref{int.eq.1} allows a meaningful interpretation of the underlying dependence structure and the univariate marginal distributions,
such a direct interpretation turns out to be difficult in a situation like \eqref{s2.eq.2}.
In addition to that,
the estimation in \eqref{s2.eq.2} is usually cumbersome since the calculation of the likelihood function can be quite involved and may require numerical integration.
Therefore, we are interested in the following naive question (Q$1$):
Can we find a $(\mu \otimes \zeta^{d})$-\,density
$h^{\ast}$ 
of the following form
\begin{equation*}
	\textrm{(Q$1$)}
	\qquad \qquad
	h^\ast (n, \yyy)
	= c \big( F(n), \GGG(\yyy) \big) \; \mathbb{P}[\{N=n\}] \; \prod_{i=1}^{d} g_i(y_i)
\end{equation*}
for $(\mu \otimes \zeta^{d})$--\,almost all $(n, \yyy) \in \mathbb{N}_0 \times \Real^d$,
and, if yes, what are the distributional properties of $h^\ast$?

\bigskip
We additionally focus on the conditional version of the distribution function $H$ of $(N,\YYY)$ in \eqref{Sklar}.
Then,
for $\mathbb{P}_N$--\,almost every $n \in \mathbb{N}_0$,
the conditional joint density $h(. | n)$ of $\YYY$ given $N=n$ equals
\begin{eqnarray}
  h(\yyy | n)
	& = & \frac{1}{\mathbb{P}[\{N=n\}]}
					 \; \left( \frac{\partial^d}{\partial \yyy} \, C \big( F(n), \GGG(\yyy) \big)
					 - \frac{\partial^d}{\partial \yyy} \, C \big( F(n-1), \GGG(\yyy) \big)\right)
					 \; \prod_{i=1}^{d} g_i(y_i) \label{CondDenD}
	\\
	& = & \frac{1}{\mathbb{P}[\{N=n\}]}
					 \; \left( \int_{(F(n-1),F(n)]}
					c \big( u, \GGG(\yyy) \big) \; \mathrm{d} \zeta(u) \right)
					 \; \prod_{i=1}^{d} g_i(y_i) \nonumber
\end{eqnarray}
for $\zeta^{d}$--\,almost all $\yyy \in \Real^d$.
In contrast to the case $(N,\YYY)$ would be absolutely continuous,
equation \eqref{CondDenD} does not lead to a practicable interpretation.
Therefore, we are also interested in the following naive question (Q$2$) related to (Q$1$):
For $\mathbb{P}_N$--\,almost every $n \in \mathbb{N}_0$,
can we find a $\zeta^{d}$-\,density
$h^{\ast} (. | n)$
conditional on $N=n$
of the following form
\begin{equation*}\label{problem2}
  \textrm{(Q$2$)}
	\qquad \qquad
	h^\ast (\yyy | n)
	= c \big( F(n), \GGG(\yyy) \big) \; \prod_{i=1}^{d} g_i(y_i)
\end{equation*}
for $\zeta^{d}$--\,almost all $ \yyy \in \Real^d$,
and, if yes, what are the distributional properties of $h^\ast(.|n)$?

\bigskip
\begin{remark}{} \label{Rem.Intro.}
If we assume
$ c (u,\GGG(\yyy)) = c (F(n),\GGG(\yyy)) $
on
$(F(n-1),F(n)] \times \Real^d$ for all $n \in \mathbb{N}_0$,
then the identities
$ h^\ast (n, \yyy) = h (n, \yyy) $ and
$ h^\ast (\yyy | n) = h (\yyy | n) $
hold for $(\mu \otimes \zeta^{d})$--\,almost all $(n, \yyy) \in \mathbb{N}_0 \times \Real^d$.
\end{remark}{}

\bigskip
In the following,
we use a generalization of the idea described in Remark \ref{Rem.Intro.} and construct a `copula'-density that is,
with respect to the first coordinate, partially constant and hence a step function.


\subsection{Transformation of the copula}
\label{sec.2.1}

For $\alpha \in (0,1]$,
we define the map
$ \lceil .\rceil_{\alpha,F}: (0,1] \to \Real$ by letting
\begin{eqnarray}\label{eq.a6}
  \lceil u\rceil_{\alpha,F}
	& := & \sum_{n \in \mathbb{N}_0} F_\alpha(n) \; \mathds{1}_{(F(n-1),F(n)]} (u)
\end{eqnarray}
where $F_{\alpha}: \mathbb{N}_0 \to \I$ is given by
$ F_\alpha (n) := (1-\alpha) \, F(n-1) + \alpha \, F(n) $,
and canonically extend $\lceil .\rceil_{\alpha,F}$
to $\I$ by putting
$ \lceil 0\rceil_{\alpha,F}
	:= 0 $.
Then, $0 \leq \lceil u\rceil_{\alpha,F} \leq 1$ for all $u \in \I$.

\bigskip
\begin{remark}{}	
We note that the identity
$ \lceil u\rceil_{\alpha,F}
	= (F_\alpha \circ F^{\leftarrow}) (u) $
holds for all $u \in (0,1)$ where $F^{\leftarrow}: (0,1) \to \Real $ denotes the {\it pseudo inverse} of $F$ given by
$ F^{\leftarrow} (u)
  := \inf \{ x \in \Real \, : \, F(x) \geq u \} $.
\end{remark}{}

\bigskip
For the copula $C$ with density function $c$,
we further define
the map $ \mathfrak{c}_{\alpha,F,C}: \I \times \I^{d} \to \Real $ by letting
\begin{equation}\label{eq.a5}
  \mathfrak{c}_{\alpha,F,C} (u,\vvv)
	:= c \big( \lceil u \rceil_{\alpha,F}, \vvv \big). 	
\end{equation}
Then $\mathfrak{c}_{\alpha,F,C}$ is positive, measurable and a $\zeta^{d+1}$-\,density:
Indeed, since
$ \partial_1 C (u, {\bf 1})
	= 1 $
holds for $\zeta$--\,almost all $u \in \I$
(where $\partial_1 C$ denotes the partial derivative of $C$ with respect to the first coordinate),
we obtain
$$
  \int_{\I \times \I^{d}} \mathfrak{c}_{\alpha,F,C} (u,\vvv) \; \mathrm{d} \zeta^{d+1} (u,\vvv)
	= \int_{\I \times \I^{d}} c \big( \lceil u \rceil_{\alpha,F}, \vvv \big)  \; \mathrm{d} \zeta^{d+1} (u,\vvv)
	= \int_{\I}  \partial_1 C (\lceil u\rceil_{\alpha,F}, {\bf 1})
				\; \mathrm{d} \zeta(u)
	= 1.
$$
The following result is immediate from Equations \eqref{eq.a6} and \eqref{eq.a5}:

\bigskip
\begin{corollary}{} \label{cor.step.fct.}
Consider $\vvv \in \I^d$.
\begin{itemize}
\item
The identity
$ \mathfrak{c}_{\alpha,F,C} (u,\vvv)
	= c(F_\alpha(n), \vvv) $
holds for all $n \in \mathbb{N}_0$ and all $u \in (F(n-1),F(n)] $.

\item
The map $\mathfrak{c}_{\alpha,F,C} (.,\vvv)$ is a positive step function.
\end{itemize}
\end{corollary}{}

\bigskip
The next example, in which we consider a pertubation of the independence copula $\Pi$
(see, e.g., \cite{dus2016, nel2006}), illustrates the construction principle
and shows that
$\mathfrak{c}_{\alpha,F,C}$ fails to be a copula density, in general:

\bigskip
\begin{example}{} \label{ExFGM1}
For $\theta \in [-1,1]$,
consider the copula $C: \I \times \I^d \to \I$
given by
$$
  C(u,\vvv)
	:= \Pi(u,\vvv) + \theta \, u(1-u) \, v_1(1-v_1) \prod_{i=2}^{d} v_i
$$
and the distribution function $F: \Real \to \I$ given by
$$
  F(x)
	:= \frac{1}{3} \; \mathds{1}_{[1,2)} (x) + \frac{2}{3} \; \mathds{1}_{[2,3)} (x) + \mathds{1}_{[3,\infty)} (x).
$$
Then we have
$	c(u,\vvv)
	= 1 + \theta \, (1-2u) (1-2v_1) $
for all $(u,\vvv) \in \I \times \I^d$
and
$$
  F_\alpha (n)
	= \begin{cases}
	    \alpha/3 	
			& n = 1; \\
			1/3 + \alpha/3
			& n = 2; \\
			2/3 + \alpha/3
			& n = 3; \\		
			1
			& n \geq 4;
    \end{cases}
	\quad \hbox{and}\quad
	\lceil u\rceil_{\alpha,F}
	= \begin{cases}
	    0 	
			& u = 0; \\
			\alpha/3
			& u \in \big( 0, \tfrac{1}{3}\big]; \\
			1/3 + \alpha/3
			& u \in \big( \tfrac{1}{3}, \tfrac{2}{3} \big]; \\		
			2/3 + \alpha/3
			& u \in \big( \tfrac{2}{3}, 1 \big].
    \end{cases}
$$
Thus, the density $\mathfrak{c}_{\alpha,F,C}$ is a step function satisfying
$$
  \mathfrak{c}_{\alpha,F,C} (u,\vvv)	
	= 1 + \theta \, (1-2v_1)
	    \begin{cases}
	    1
			& u = 0; \\
			\frac{3-2\alpha}{3}
			& u \in \big( 0, \tfrac{1}{3}\big]; \\
			\frac{1-2\alpha}{3}
			& u \in \big( \tfrac{1}{3}, \tfrac{2}{3} \big]; \\		
			\frac{-1-2\alpha}{3}
			& u \in \big( \tfrac{2}{3}, 1 \big],
    \end{cases}
$$
for all $\vvv \in \I^d$.
Since, for every $\alpha \neq 1/2$ and $\theta \neq 0$ and every $v_1 \in (0,1)$,
$$
  \int_{\I \times [0,v_1] \times \I^{d-1}}
	\mathfrak{c}_{\alpha,F,C} (s,\ttt)
	\; \mathrm{d} \zeta^{d+1}(s,\ttt)
	= v_1 + \theta \, v_1 \, (1-v_1) \left( \frac{1-2\alpha}{3} \right)
	\neq v_1,
$$
$\mathfrak{c}_{\alpha,F,C}$ fails to be a copula density.
\end{example}{}

\bigskip
Obviously,
if $\mathfrak{c}_{\alpha,F,C}$ fails to be a copula density,
then $\mathfrak{c}_{\alpha,F,C} \neq c$.
Note that the converse implication is not true, in general.
For instance, take $\alpha = 1/2$ and $\theta \neq 0$ in Example \ref{ExFGM1}.
Then, it is straightforward to show that $\mathfrak{c}_{\alpha,F,C}$ is a copula density but fails to coincide with $c$.
Nevertheless, there exist copulas satisfying $\mathfrak{c}_{\alpha,F,C} = c$.

\bigskip
\begin{example}{} \label{ExPi}
Consider a $d$-dimensional copula $A$ with $\zeta^d$-\,density $a$ and define the map
$C: \I \times \I^d \to \I $ by letting
$ C(u,\vvv)
	:= u \, A (\vvv) $.
Then, by \cite[Theorem 6.6.3]{sas1983},
$C$ is a $(d+1)$-dimensional copula with $\zeta^{d+1}$-\,density
$c$ satisfying
$ \mathfrak{c}_{\alpha,F,C} (u,\vvv)
	= c \big( \lceil u \rceil_{\alpha,F}, \vvv \big)
	= a (\vvv)
	= c(u,\vvv) $
for all $(u,\vvv) \in \I \times \I^d$.
\end{example}{}

\bigskip
Since $\mathfrak{c}_{\alpha,F,C}$ is a $\zeta^{d+1}$-\,density,
the map $\mathfrak{C}_{\alpha,F,C}: \I \times \I^{d} \to \Real$ given by
\begin{equation}
  \mathfrak{C}_{\alpha,F,C} (u,\vvv)
	:= \int_{[0,u] \times [{\bf 0},\vvv]} \mathfrak{c}_{\alpha,F,C} (s,\ttt) \; \mathrm{d} \zeta^{d+1}(s,\ttt)
\end{equation}
is a distribution function on $\I^{d+1}$ following an absolutely continuous distribution.
In the next lemma
we gather some properties of $\mathfrak{C}_{\alpha,F,C}$
that turn out to be quite useful.

\bigskip
\begin{lemma}{} \label{densityprop}
The identity
\begin{eqnarray*}
  \mathfrak{C}_{\alpha,F,C} (u,\vvv)
	& = & \sum_{k=0}^{n-1} \partial_1 C (F_{\alpha}(k),\vvv) \, \mathbb{P}[\{N=k\}]
				+ \partial_1 C (F_{\alpha}(n),\vvv) \, \big( u-F(n-1) \big)
\end{eqnarray*}
holds for all $n \in \mathbb{N}_0$
and all $(u,\vvv) \in (F(n-1),F(n)] \times \I^d$.
In particular, we have
\begin{equation*}
  \mathfrak{C}_{\alpha,F,C} (u,{\bf 1})= u
	\quad\hbox{and}\quad
  \mathfrak{C}_{\alpha,F,C} (1,\vvv)
  = E \big[ \partial_1 C \big(F_\alpha(N), \vvv\big) \big]				
\end{equation*}
for all $u \in \I$ and all $\vvv \in \I^d$.
\end{lemma}{}

\bigskip
\begin{proof}{}
For every $n \in \mathbb{N}_0$ and every $u \in (F(n-1),F(n)]$,
Corollary \ref{cor.step.fct.} yields
\begin{eqnarray*}
  \int_{[0,u]} \mathfrak{c}_{\alpha,F,C} (s,\vvv) \; \mathrm{d} \zeta(s)
	& = & \int_{[0,u]} c \big( \lceil s \rceil_{\alpha,F}, \vvv \big) \; \mathrm{d} \zeta(s)
	\\
	& = & \sum_{k=0}^{n-1} \int_{(F(k-1),F(k)]} c (F_{\alpha}(k),\vvv) \; \mathrm{d} \zeta(s)
				+ \int_{(F(n-1),u]} c (F_{\alpha}(n),\vvv) \; \mathrm{d} \zeta(s)
	\\
	& = & \sum_{k=0}^{n-1} c (F_{\alpha}(k),\vvv) \, \big( F(k)-F(k-1) \big)
				+ c (F_{\alpha}(n),\vvv) \, \big( u-F(n-1) \big)
	\\
	& = & \sum_{k=0}^{n-1} c (F_{\alpha}(k),\vvv) \, \mathbb{P}[\{N=k\}]
				+ c (F_{\alpha}(n),\vvv) \, \big( u-F(n-1) \big)
\end{eqnarray*}
for $\zeta^d$--\,almost all $\vvv \in \I^d$.
This proves the assertion.
\end{proof}{}

\bigskip
From a probabilistic viewpoint,
if a random vector $(U,{\bf V})$ is distributed according to $\mathfrak{C}_{\alpha,F,C}$,
then it follows from Lemma \ref{densityprop}
that $U$ is uniformly distributed, i.e.
$$\mathbb{P} [\{U\le u\}]= u$$ for all $u\in\I$,
and that the identity
$$
  \mathbb{P} [\{ {\bf V} \le \vvv \}]
	= E \big[ \partial_1 C \big(F_\alpha(N), \vvv\big) \big]
$$
holds for all $\vvv \in \I^d$.

\bigskip
We illustrate the construction principle by completing Example \ref{ExFGM1}.

\bigskip
\begin{example}{}\label{ex.2.7}
For $\theta \in [-1,1]$,
consider the copula $C$
and the distribution function $F$
discussed in Example \ref{ExFGM1}.
Then
\begin{eqnarray*}
	\mathfrak{C}_{\alpha,F,C} (u,\vvv)
	& = & \Pi(u,\vvv) + \theta \, v_1 (1-v_1) \prod_{i=2}^{d} v_i
	    \begin{cases}
	    0
			& u = 0 \\
			\big( \frac{3-2\alpha}{3} \big) \, u
			& u \in \big( 0, \tfrac{1}{3}\big] \\
			\frac{2}{9} + \; \big( \frac{1-2\alpha}{3} \big) \, u
			& u \in \big( \tfrac{1}{3}, \tfrac{2}{3} \big] \\		
			\frac{6}{9} + \; \big( \frac{-1-2\alpha}{3} \big) \, u
			& u \in \big( \tfrac{2}{3}, 1 \big]
    \end{cases}
\end{eqnarray*}
for all $\vvv \in \I^d$
and hence
$ \mathfrak{C}_{\alpha,F,C} (u,{\bf 1}) = u $
for every $u \in \I$ as well as
$$
  \mathfrak{C}_{\alpha,F,C} (1,\vvv)
	= E \big[ \partial_1 C \big(F_\alpha(N), \vvv\big) \big]
	= \prod_{i=1}^{d} v_i + \theta \, v_1 (1-v_1) \prod_{i=2}^{d} v_i \left( \frac{1-2\alpha}{3} \right)
	\in \I
$$
for every $\vvv \in \I^d$.
Since, for every $\alpha \neq 1/2$ and $\theta \neq 0$ and every $v_1 \in (0,1)$
$$
  \mathfrak{C}_{\alpha,F,C} \big(1,(v_1,{\bf 1})\big)
	= v_1 + \theta \, v_1 (1-v_1) \left( \frac{1-2\alpha}{3} \right)
	\neq v_1,
$$
$\mathfrak{C}_{\alpha,F,C}$ fails to be a copula.
\end{example}{}

\bigskip
Choosing $\theta=0$ in the previous example yields
$C=\Pi$ and hence $\mathfrak{C}_{\alpha,F,C}=\Pi=C$.
A more general result is given by the following example which extends Example \ref{ExPi}.

\bigskip
\begin{example}{}
Consider a $d$-dimensional copula $A$ with $\zeta^d$-\,density and the copula
$C: \I \times \I^d \to \I $ given by
$ C(u,\vvv)
	:= u \, A (\vvv) $.
Then,
$ \mathfrak{C}_{\alpha,F,C} = C $.
In particular, $ \mathfrak{C}_{\alpha,F,\Pi} = \Pi $.
\end{example}{}


\subsection{Transformation of the random vector}\label{sec.2.2}

Although the copula transformation $\mathfrak{C}_{\alpha,F,C}$ of $C$ fails to be a copula, in general,
it is a distribution function on $\I^{d+1}$ whose first coordinate is distributed uniformly.
In this subsection we will use this copula transformation in combination with Sklar's theorem to construct a distribution function that helps answering questions (Q$1$) and (Q$2$).

\bigskip
For $\alpha \in (0,1]$,
the copula $C$ with density function $c$,
the discrete distribution function $F$ and absolutely continuous distribution functions $G_1, \dots, G_d$,
we define the function $\mathfrak{H}_{\alpha,F,\GGG,C}: \Real \times \Real^d \to \I$ by letting
\begin{eqnarray*}
  \mathfrak{H}_{\alpha,F,\GGG,C} (x,\yyy)
	& := & \mathfrak{C}_{\alpha,F,C} \big( F(x), \GGG(\yyy) \big).
\end{eqnarray*}
Then it is straightforward to verify that $\mathfrak{H}_{\alpha,F,\GGG,C}$ is a distribution function
satisfying
\begin{eqnarray*}
\lim_{\ttt \to {\bf \infty}} \mathfrak{H}_{\alpha,F,\GGG,C} (x,\ttt)
	= F(x)
	\quad\hbox{and}\quad
	\lim_{s \to \infty} \mathfrak{H}_{\alpha,F,\GGG,C} (s,\yyy)
	= E \big[ \partial_1 C \big( F_\alpha (N), \GGG(\yyy) \big) \big].
\end{eqnarray*}

\bigskip
The next result is immediate from Corollary \ref{cor.step.fct.} and solves question (Q$1$).

\bigskip
\begin{theorem}{} \label{Q1}
The $(\mu \otimes \zeta^{d})$-\,density $\mathfrak{h}_{\alpha,F,\GGG,C}$ of $\mathfrak{H}_{\alpha,F,\GGG,C}$ satisfies
$$
  \mathfrak{h}_{\alpha,F,\GGG,C} (n,\yyy)
	= c \big( F_\alpha (n), \GGG(\yyy) \big)
				\, \mathbb{P} [\{N=n\}]
				\; \prod_{i=1}^{d} g_i(y_i)
$$
for $(\mu \otimes \zeta^{d})$-\,almost all $(n, \yyy) \in \mathbb{N}_0 \times \Real^d$.
\end{theorem}{}

\bigskip
From a probabilistic viewpoint,
if a random vector $(M,{\bf T})$ is distributed according to $\mathfrak{H}_{\alpha,F,\GGG,C}$,
then Lemma \ref{densityprop} shows that $M$ is a random variable whose distribution function $F_M$ equals
$$
  F_M	= F
$$
and that the distribution function $F_{{\bf T}}$ of ${\bf T}$ satisfies
$$
  F_{{\bf T}}
	= E \big[ \partial_1 C \big( F_\alpha (N), \GGG(.) \big) \big].
$$
Note that $F_{T_i} \neq G_i$, $i \in \{1,\dots,d\}$, in general; see Example \ref{ex.2.7}.
Now, we answer question (Q$2$).

\bigskip
\begin{theorem}{} \label{Q2}
Consider a random vector $(M,{\bf T})$ distributed according to $\mathfrak{H}_{\alpha,F,\GGG,C}$.
Then, for $\mathbb{P}_M$--\,almost every $n \in \mathbb{N}_0$,
the conditional joint density $\mathfrak{h}_{\alpha,F,\GGG,C} (. | n)$ of ${\bf T}$ given $M=n$
satisfies
\begin{equation}\label{eq.a4}
  \mathfrak{h}_{\alpha,F,\GGG,C} (\ttt | n)
	= c \big( F_\alpha (n), \GGG(\ttt) \big) \; \prod_{i=1}^{d} g_i(t_i)
\end{equation}
for $\zeta^{d}$--\,almost all $ \ttt \in \Real^d$.
\end{theorem}{}

\bigskip
\begin{remark}{}
In the special case $\alpha=1$,
the results in Theorems \ref{Q1} and \ref{Q2} reduce to
$$
  \mathfrak{h}_{\alpha,F,\GGG,C} (n,\yyy)
	= c \big( F(n), \GGG(\yyy) \big)
				\, \mathbb{P} [\{N=n\}]
				\; \prod_{i=1}^{d} g_i(y_i)
$$
and
\begin{equation*}
  \mathfrak{h}_{\alpha,F,\GGG,C} (\yyy | n)
	= c \big( F (n), \GGG(\yyy) \big) \; \prod_{i=1}^{d} g_i(y_i)
\end{equation*}
for $(\mu \otimes \zeta^{d})$--\,almost all $(n, \yyy) \in \mathbb{N}_0 \times \Real^d$.
\end{remark}


\section{Numerical Analysis}
\label{Sec.NA.}

In this section, we perform a numerical study illustrating the impact of the copula transformation method suggested in Section \ref{sec.2.1}
by measuring the distance between the copula $C$ and its transformed version $\mathfrak{C}_{\alpha,F,C}$.

\bigskip
As distance measure we use the Kullback-Leibler divergence (KL divergence for short);
see, e.g., \citet{Kullback1951, Kullback1997}.
For two $k$-dimensional joint distribution functions $P$ and $Q$ having $p$ and $q$ as probability density functions,
the KL divergence from $P$ to $Q$ is defined as
\[
  D(P, Q)
	:= \int_{\Real^k} p(\xxx) \, \log \left(\frac{p(\xxx)}{q(\xxx)}\right) \; \mathrm{d} \zeta^k (\xxx).
\]
We note that $D(P, Q)\ge 0$ where equality holds if and only if $P=Q$.
In addition, since $D(P, Q) \neq D(Q, P)$, KL divergence fails to be symmetric, in general.

\bigskip
In the first part of our numerical analysis,
we consider bivariate copulas $C_\theta$ from various parametric copula families
(Gaussian, Student $t$, Clayton and Gumbel) and put
\[
  P = C_\theta
	\qquad \hbox{and} \qquad
	Q = \mathfrak{C}_{\alpha,F,C_\theta}
\]
where $F$ denotes a Poisson distribution with mean $\zeta$.
The parameter $\theta$ is choosen in such way that it corresponds to a certain value of bivariate Kendall's tau and hence
indicates the degree of dependence represented by $C_\theta$.
We measure the KL divergence from $P$ to $Q$ under various scenarios which are combinations of
$\alpha=0.25, 0.5, 0.75, 1.0$,
$\zeta=0.1, 0.5, 1.0, 5.0, 10.0$
and $\theta$ given in Table \ref{tb.theta}.
\begin{table}[h]
\centering
\begin{tabular}{|c|c|c|c|c|c|c|c|c|}
\multicolumn{5}{l}{}\\
  \hline
  \multicolumn{2}{|l|}{Kendall's tau}& -0.8	&-0.3	&-0.1	&0	&0.1	&0.3 	&0.8 \\
  \hline\hline
   \multirow{4}{*}{$\theta$}
&	Gaussian / Student t	&	-0.951 	&	-0.454 	&	-0.156 	&	0.000 	&	0.156 	&	0.454 	&	0.951 	\\ \cline{2-9}
&	Clayton	&	-		&		-	&	-		&	0.000 	&	0.222 	&	0.857 	&	8.000 	\\ \cline{2-9}
&	Gumbel	&	-		&		-	&	-		&	1.000 	&	1.111 	&	1.429 	&	5.000 	\\ \cline{2-9}
 \hline
\end{tabular}
\caption{Copula parameter $\theta$ corresponding to bivariate Kendall's tau  }\label{tb.theta}
\end{table}

For each scenario, we further calculate the values of Spearman's rho
\[
  \rho(P)
	\qquad \hbox{and} \qquad
	\rho(Q)
\]
to illustrate how the dependence structure of the copula changes with the proposed transformation;
recall that Spearman's rho of a bivariate distribution function $H$ with marginals $F$ and $G$ is defined as
\[
\rho(H):=12\int_{\Real^2} F(x)G(y) \; \mathrm{d} H(x,y) - 3
\]

In the second part of the numerical analysis, we leave the bivariate setting and consider 3-dimensional copulas.
Here, we restrict ourselves to positive dependence (symmetric Gaussian and Clayton copulas) whereas the parameter $\theta$ is chosen in such a way that it corresponds to a certain value of the bivariate Kendall's tau as in Table \ref{tb.theta}.


\bigskip
The results of the numerical analysis are summarized in Table \ref{tb.kld.1} to
\ref{tb.kld.d3.2}:
\begin{bulllist}
\item Table \ref{tb.kld.1} to \ref{tb.kld.4}: KL divergence for Gaussian, Student $t$, Clayton and Gumbel copula in dimension $2$.
\item Table \ref{tb.rho.1} to \ref{tb.rho.4}: Spearmans's rho for Gaussian, Student $t$, Clayton and Gumbel copula in dimension $2$.
\item Table \ref{tb.kld.d3.1} and \ref{tb.kld.d3.2}: KL divergence for the $3$-dimensional Gaussian and Clayton copula.
\end{bulllist}

\bigskip
From Table \ref{tb.kld.1} and \ref{tb.kld.2}, one may observe that,
for each $\alpha$ and $\zeta$, the values of KL divergence are symmetric about $0$ with respect to $\theta$ which is due to the fact that the density functions of two Gaussian copulas (t-copulas) having parameters of opposite signs, $\pm \rho$, are reflection of each other over the horizon line $x=0.5$.
Similarly, from Table \ref{tb.rho.1} and \ref{tb.rho.2}, one may also observe that,
for each $\alpha$ and $\zeta$, the values $\rho(P)$ and $\rho(Q)$ are symmetric about $0$.

\bigskip
As expected from the definition of the transformation $\mathfrak{C}_{\alpha,F,C_\theta}$,
we observe that copulas with weaker dependence tend to have smaller KL divergence and smaller discrepancy between $\rho(P)$ and $\rho(Q)$.
We also observe that for each $\theta$, as $\zeta$ increases, KL divergence decreases and the discrepancy is diminished,
which is also an expected result from the definition of the transformation $\mathfrak{C}_{\alpha,F,C_\theta}$.
However, one interesting phenomenon discovered in this numerical analysis is that, around $\alpha=0.5$,
we observe the smallest KL divergence for each combination of $\zeta$ and $\theta$.
Finally, as can be shown in Table \ref{tb.kld.d3.1} and \ref{tb.kld.d3.2}, we note that 3-dimensional copulas show similar patterns as in 2-dimensional copulas.




\section{Application to Collective Risk Model}
\label{Sec.CRM.}

In this section,
we apply the proposed copula transformation method to the collective risk model (CRM, for short).
In the CRM, the \emph{aggregate severity} in insurance portfolio is modelled as the random sum of individual severities.
Specifically, for the nonnegative integer valued random variable $N$ and the positive random variables $Y_j$,
the aggregate severity $S$ is defined by
   \[
   S:= \begin{cases}
     \sum\limits_{j=1}^{N} Y_j, & \hbox{if }\, N=n\in\mathbb{N};\\
     0,												 & \hbox{if }\, N=0.\\
   \end{cases}
   \]
Note that the aggregate severity can be expressed as
$S=MN$
where $M$ is the \emph{average severity} given by
   \[
   M:= \begin{cases}
    \dfrac{1}{N} \sum\limits_{j=1}^{N} Y_j, & \hbox{if }\, N=n\in\mathbb{N};\\
     0,																		   & \hbox{if }\, N=0.\\
   \end{cases}
   \]

We first review two CRMs in the insurance literature.
The first model that we will revisit is so called the {\it two part} CRM where dependence between frequency and severity is induced by
using the frequency as an explanatory variable of the severities; see, e.g, \citet{Frees2, Peng, Garrido, Ahn2008}.
As a result, in this model the distribution of the aggregate severity can be easily determined.
However, it is known that the dependence structure in the two part model is quite limited;
see, e.g., \citet{liu2017collective, shi2020regression}.
The second model that we will revisit is the {\it copula-based} CRM, which can cover the full spectrum of dependencies by describing the dependence for the frequency and the individual severities based on copula function \citep{cossette2019collective, oh2020copulaSAJ}.
Note that the description of the aggregate severity, which is the main concern of insurance industry, under the copula-based CRM can be inconvenient as will be explained below.
Finally, we apply the proposed copula transformation to the copula based CRM which enables convenient handling of the distribution of the aggregate severity, and we provide an example where the linkage between the two CRMs is demonstrated.


\subsection{Two part CRM for frequency and aggregate severity}

\begin{model}{} \label{mod.a1}
The \emph{two part CRM for frequency and aggregate severity}
$$\left(N, M \right)$$ is defined
within the framework of the exponential dispersion family (EDF) as follows
(see \cite{Frees2, Garrido}):
\begin{romlist}
\item
We specify the frequency component $N$ as
\begin{equation}\label{eq.ahn1}
  N \sim F
\end{equation}
where $F$ can be any discrete distribution function on $\mathbb{N}_0$.

\item
We specify the conditional distribution of the average severity conditional on the number of claims
$N=n \in \mathbb{N}$ as
\begin{equation}\label{eq.ahn2}
  M\big\vert N=n\iid {\rm ED}\left(\mu_n, \sigma^2_n\right)
\end{equation}
where ${\rm ED}\left(\mu_n, \sigma^2_n\right)$ is the reproductive exponential dispersion model with mean $\mu_n$ and dispersion parameter $\sigma_n^2$; see, e.g., \citet{jorgensen1987exponential, jorgensen1997theory}.
Here, the mean parameter $\mu_n$ is implicitly given by 
$\eta_1\left(\mu_n\right)= \beta_1+\psi_1(n)$ for properly chosen function $\psi_1$ and link function $\eta_1$, and the dispersion parameter is given by
\begin{equation}\label{lambda}
    \sigma_n^2:=\frac{\sigma_0^2}{n}.
\end{equation}
\end{romlist}
\end{model}{}

\bigskip
The choice of the dispersion parameter in \eqref{lambda} can be justified by the following distributional assumption on the individual severities
\begin{equation}\label{eq.ahn2a}
  Y_1, \dots, Y_n|N=n \iid {\rm ED}\left(\mu_n, \sigma_0^2\right)
	\qquad n\in\mathbb{N}
\end{equation}
which implies \eqref{eq.ahn2} by the convolutionary property of EDF.
As a result, one may replace the description of the average severity in Model \ref{mod.a1} with the description of the individual severities in \eqref{eq.ahn2a}. We call such model as the {\it two part CRM} for frequency and the individual severities.
However, while convenient in many ways,
the conditional independence assumption in \eqref{eq.ahn2a}
is a rather restrictive dependence assumption of frequency and individual severities
as pointed out in \cite{liu2017collective} and \cite{shi2020regression}.
Therefore, the two part CRM with the functional form of the dispersion parameter in \eqref{lambda} can accommodate only restrictive dependence structures of frequency and individual severities.

\bigskip
Alternatively,
one may choose a more complicated functional form of $\sigma_n$ as mentioned in \cite{lee2019investigating} to consider more general dependence structures in the CRM.
Depending on the purpose of the data analysis,
one may use, for instance, advanced regression modeling strategies such as non-parametric regression or additive modeling; see, e.g., \citet{Hastie, Faraway}.
However, 
in such a case
the important linkage between the average severity in \eqref{eq.ahn2} and the individual severities in \eqref{eq.ahn2a} is violated,
in general.



\subsection{The copula-based CRM for the frequency and the individual severities}

We now consider the copula-based CRM for the frequency and the individual severities discussed in \cite{cossette2019collective} and \cite{oh2020copulaSAJ}
where a wider variety of dependence structures is possible depending on the particular choice of the used copula family. 

\bigskip
\begin{model}{}\label{mod.a2}
The \emph{copula-based CRM} for frequency and individual severities
$$\left(N, Y_1, Y_2, \dots, Y_N \right)$$ is defined as follows:
\begin{romlist}
\item
We specify the frequency component $N$ as
\begin{equation}\label{eq.ahn11}
  N \sim F
\end{equation}
where $F$ can be any discrete distribution function on $\mathbb{N}_0$; compare \eqref{eq.ahn1}.
\item
We specify the conditional distribution of the vector of individual severities $(Y_1, \cdots, Y_n)$ conditional on the number of claims $N=n\in\mathbb{N}$ as
\begin{equation}\label{eq.ahn20}
  (Y_1, \dots, Y_n)\big\vert N=n \sim W_{(n)}
\end{equation}
for some distribution function $W_{(n)}$ given by
\begin{equation*}
  W_{(n)} (y_1, \dots, y_n)
	:= \frac{C_{(n)} \big( F(n), G(y_1), \dots, G(y_n) \big) - C_{(n)} \big( F(n-1), G(y_1), \dots, G(y_n) \big)}{\mathbb{P} [\{N=n\}]}
        \end{equation*}
with $C_{(n)}$ being an absolutely continuous $(n+1)$-dimensional copula.
\end{romlist}
\end{model}{}

\bigskip
In this model,
the density function of $\left(Y_1, Y_2, \cdots, Y_n \right)$ at point $\left(y_1, y_2, \cdots, y_n \right)$
conditional on $N=n\in\mathbb{N}$ satisfies
\begin{equation}\label{eq.ahn22}
  \frac{\partial^n}{\partial y_1 \dots \partial y_n} \; W_{(n)}(y_1, \dots, y_n)
  = \left( \int_{(F(n-1),F(n)]} c_{(n)} \big( u, G(y_1), \dots, G(y_n) \big) \; \mathrm{d} \zeta(u) \right)
	  \; \frac{\prod_{i=1}^{n}g(y_i)}{\mathbb{P} [\{N=n\}]}.
\end{equation}
Hence, loosely speaking, the copula-based CRM can be understood as
\begin{equation}\label{eq.ahn30}
(N, Y_1, \dots, Y_N)\sim C_{(N)}(F, G, \dots, G).
\end{equation}
We refer to \cite{cossette2019collective} and \cite{oh2020copulaSAJ} for the natural linkage between \eqref{eq.ahn22} and the density function in \eqref{eq.ahn30}.

\bigskip
The copula-based CRM allows modeling both, the dependence among the individual severities and the dependence between frequency and severities.
Hence, in terms of the dependence structure, the copula-based CRM provides a wider range of dependence than the two part CRM.
But while the main interest of insurance industry lies in the aggregate severity $S$,
the copula-based CRM does not allow an analytical interpretation of $S$, in general.
For an analysis of $S$ one may consider to use the conditional distribution in \eqref{eq.ahn20}.
However, mainly due to the non-continuous nature of the frequency $N$,
an analytical interpretation of the conditional distribution in \eqref{eq.ahn20} or equivalently in \eqref{eq.ahn22} is difficult, in general.


\subsection{The transformed copula-based CRM for the frequency and the individual severities}

In the following, we provide an example where the proposed copula transformation method allows an analytical interpretation of the conditional dependence among severities in \eqref{eq.ahn20} as well as the dependence between frequency and individual severities.
The following model is a modification of the previous copula-based CRM where the copula $C$ is replaced by its transformed version $\mathfrak{C}_{\alpha,F,C_{(n)}}$.

\bigskip
\begin{model}{}\label{mod.1}
For $\alpha\in(0,1)$,
the \emph{transformed copula-based CRM} for
$$\left(N, Y_1, Y_2, \dots, Y_N \right)$$ is defined as follows:
\begin{romlist}
\item
We specify the frequency component $N$ as
\begin{equation}
  N \sim F
\end{equation}
where $F$ can be any discrete distribution function on $\mathbb{N}_0$; compare \eqref{eq.ahn1} and \eqref{eq.ahn11}.

\item
We specify the conditional distribution of the vector of individual severities $(Y_1, \dots, Y_n)$ conditional on the number of claims $N=n\in\mathbb{N}$ as
        \begin{equation}\label{eq.ahn40}
               (Y_1, \dots, Y_n)\big\vert N=n \sim W_{(n)}^\ast
        \end{equation}
        for some distribution function $W_{(n)}^\ast$ given by
        \begin{equation}\label{eq.ahn41}
        W_{(n)}^\ast\left(y_1, \dots, y_n \right)
				:=\frac{\mathfrak{C}_{\alpha,F,C_{(n)}} \big(F(n), G(y_1), \dots, G(y_n) \big)-\mathfrak{C}_{\alpha,F,C_{(n)}} \big( F(n-1), G(y_1), \dots, G(y_n) \big)}{\mathbb{P} [\{N=n\}]}
        \end{equation}
        with $C_{(n)}$ being an absolutely continuous $(n+1)$-dimensional copula.
\end{romlist}
\end{model}{}

\bigskip
Note that, due to Theorem \ref{Q2}, the conditional distribution function in \eqref{eq.ahn40} can be represented as
\[
  W_{(n)}^*(y_1, \dots, y_n)
  = \partial_1 C_{(n)} \big( F_\alpha (n), G(y_1), \dots, G(y_n) \big)
\]
and the corresponding density function at $\left(y_1, y_2, \cdots, y_n \right)$
conditional on $N=n\in\mathbb{N}$ satisfies
\begin{equation}\label{eq.ahn42}
  \frac{\partial^n}{\partial y_1 \dots \partial y_n} \; W_{(n)}^\ast (y_1, \dots, y_n)
	= c_{(n)} \big( F_\alpha(n), G(y_1), \cdots, G(y_n) \big) \prod_{i=1}^{n}g(y_i)
\end{equation}
Similar to the case of copula-based CRM,
the transformed copula-based CRM can be understood as
\begin{equation}\label{eq.ahn50}
(N, Y_1, \dots, Y_N)\sim \mathfrak{C}_{\alpha,F,C_{(N)}}(F, G, \dots, G)
\end{equation}


\bigskip
Now we are ready to provide an example which shows a link between two part CRM and copula-based CRM.
First, we define the following matrices for $\rho_1, \rho_2\in[-1,1]$:
\begin{romlist}
\item
For $k \in \mathbb{N}$ and $l \in \{1,2\}$,
define the $k\times k$ matrix $\boldsymbol{\Sigma}_{\rho_2}^{[k,l]}$ by letting
  \begin{equation*}
  \left[\boldsymbol{\Sigma}_{\rho_2}^{[k,1]}\right]_{i,j}:=\begin{cases}
    1, 			 &\hbox{if}\quad i=j;\\
    \rho_2,   &\hbox{if}\quad i\neq j;\\
  \end{cases}
  \end{equation*}
and
  \begin{equation*}
  \left[\boldsymbol{\Sigma}_{\rho_2}^{[k,2]}\right]_{i,j}:=\begin{cases}
    1, &\hbox{if}\quad i=j;\\
    \rho_2^{\left\vert i-j\right\vert}, &\hbox{if}\quad i\neq j.\\
  \end{cases}
  \end{equation*}

\item
For $k \in \mathbb{N}$ and $l \in \{1,2\}$,
define the $(k+1)\times(k+1)$ matrix $\boldsymbol{\Sigma}_{\rho_1,\rho_2}^{[k,l]}$ 
by letting
  \begin{equation*}
    \boldsymbol{\Sigma}_{\rho_1,\rho_2}^{[k,l]}:=\begin{cases}
    \left(
    \begin{array}{cc}
      1 & \rho_1 \left({\bf 1}_k\right)^{\mathrm T}\\
      \rho_1 {\bf 1}_k & \boldsymbol{\Sigma}_{\rho_2}^{[k,l]}\\
    \end{array}
    \right), & k \in \mathbb{N}; \\
     1, & k=0;\\
    \end{cases}
  \end{equation*}
where ${\bf 1}_k$ is a column vector with entries $1$ of length $k$.
\end{romlist}
As shown in \cite{oh2020copulaSAJ}, the condition
\begin{equation} \label{oh.22}
(\rho_1, \rho_2)\in\left\{ (\rho_1, \rho_2)\in(-1,1)^2 \, \big\vert \, \rho_1^2< \rho_2< 1 \right\}
\end{equation}
implies positive definiteness of the matrix $\boldsymbol{\Sigma}_{\rho_1,\rho_2}^{[k,1]}$ for every $k\in\mathbb{N}_0$.
Similarly, using the well known result on the Schur complement of a block matrix \citep{haynsworth1968schur},
 the matrix $\boldsymbol{\Sigma}_{\rho_1,\rho_2}^{[k,2]}$ is positive definite
for any $\rho_1, \rho_2\in(-1,1)$ satisfying
\begin{equation} \label{oh.23}
1-\rho_1^2\left( k\left( 1-\rho_2 \right)+2\rho_2\right)\left( 1-\rho_2\right)>0.
\end{equation}
We denote by
$ C \big( \cdot ;\boldsymbol{\Sigma}_{\rho_1,\rho_2}^{[n,l]} \big) $
the $(n+1)$-dimensional Gaussian copula
with correlation matrix $\boldsymbol{\Sigma}_{\rho_1,\rho_2}^{[n,l]}$
and by $c\big(\cdot ;\boldsymbol{\Sigma}_{\rho_1,\rho_2}^{[n,l]}\big)$ its density.

\bigskip
In the sequel,
we consider Model \ref{mod.1} for $\left(N, Y_1, Y_2, \cdots, Y_N \right)$
assuming a symmetric dependence structure for the individual severities:

\bigskip
\begin{assumption}{}\label{assum.1}
\begin{bulllist}
\item $\rho_1$ and $\rho_2$ satisfy condition \eqref{oh.22}.
\item
$C_{(n)}$ is an $(n+1)$-dimensional Gaussian copula
with correlation matrix $\boldsymbol{\Sigma}_{\rho_1,\rho_2}^{[n,1]}$.

\item
$G$ is a normal distribution with mean $\xi$ and variance $\sigma^2$.
\end{bulllist}
\end{assumption}

\bigskip
In this case, the conditional density function of $\left(Y_1, Y_2, \cdots, Y_n \right)$ conditional on $N=n \in \mathbb{N}$ satisfies
\begin{equation}\label{eq.ahn62}
  \frac{\partial^n}{\partial y_1 \dots \partial y_n} \; W_{(n)}^*(y_1, \cdots, y_n)
  = c_{(n)} \left(F_\alpha(n), \Phi_{\xi, \sigma^2}(y_1), \dots, \Phi_{\xi, \sigma^2}(y_n);\boldsymbol{\Sigma}_{\rho_1,\rho_2}^{[n,1]}\right)\prod_{i=1}^{n}\phi_{\xi, \sigma^2}(y_i)
\end{equation}
and part i of Lemma \ref{lemma.1} shows that, for $n\in\mathbb{N}$,
$\left(Y_1, \cdots, Y_n \right)$ conditional on $N=n$ follows a multivariate normal distribution
with mean $\big( \xi + \sigma\rho_1\Phi^{-1}_{0,1} (F_\alpha(n)) \big)\boldsymbol{1}_n$ and covariance matrix
$$
    \sigma^2\left( \boldsymbol{\Sigma}_{\rho_2}^{[n,1]}-\rho_1^2 \boldsymbol{J}_{n\times n} \right)
$$
where $\boldsymbol{J}_{n\times n}$ denotes an $n \times n$ matrix with entries $1$.
By the convolutionary property of the multivariate normal distribution, we then obtain
\[
\frac{Y_1+\cdots+Y_n}{n} \,\Big|\, N=n \sim {\rm N}\left( \mu_n, \sigma_n^2\right)
\]
 where
\[
\mu_n= \xi + \sigma\rho_1\Phi_{0,1}^{-1}(F_\alpha(n))
\quad\hbox{and}\quad
\sigma_n^2=\frac{1}{n}\sigma^2\big( (n-1)\rho_2 - n\rho_1^2+1\big)
\]
As a result, the distribution of $S$ can be expressed in closed form which in turn allows a closed form expression for subsequent statistics of $S$.
Specifically, for $s\ge 0$, we have
\[
\begin{aligned}
\P{S\le s}&=\sum\limits_{n=0}^{\infty}\P{S\le s\} | \{N=n} \, \P{ N=n}\\
&=F(0)+\sum\limits_{n=1}^{\infty}\Phi\left( \frac{s/n -\mu_n}{\sigma_n}\right) \, \P{ N=n}
\end{aligned}
\]
Additionally, we obtain
$$
  \E{S}
	=\E{N \, \E{M | N}}
  =\sum\limits_{n=1}^{\infty} n \mu_n \, \P{ N=n}
$$
and
\[
\begin{aligned}
\Var{S}&=\E{N^2 \, \Var{M|N}}+\Var{N \, \E{M|N}}\\
&=\sum\limits_{n=1}^{\infty} \left( n^2\sigma_n^2 +n^2\mu_n^2 \right) \, \P{ N=n}
-\left( \sum\limits_{n=1}^{\infty} n\mu_n  \, \P{ N=n}    \right)^2.
\end{aligned}
\]

\bigskip
Note that, while the dependence structure of the conditional severities in \eqref{eq.ahn2a} under two part CRM is restricted to conditional independence,
the dependence structure of the conditional severities in \eqref{eq.ahn41} under transformed copula-based CRM allows more general dependence structures.
In Lemma \ref{lemma.1}, we also provide the condition where
two part CRM and transformed copula-based CRM are equivalent.

\bigskip
Finally,
we consider Model \ref{mod.1} for $\left(N, Y_1, Y_2, \cdots, Y_N \right)$
assuming an autoregressive dependence structure for the individual severities.

\bigskip
\begin{assumption}{}\label{assum.2}
\begin{bulllist}
\item $F$ is a discrete distribution function with finite support on $\mathbb{N}_0$ having $\kappa_0\in\mathbb{N}$ as the essential supremum of $F$.
\item $k=\kappa_0$ satisfies \eqref{oh.23}.
\item
$C_{(n)}$ is an $(n+1)$-dimensional Gaussian copula
with correlation matrix $\boldsymbol{\Sigma}_{\rho_1,\rho_2}^{[n,2]}$ for $n\le \kappa_0$.

\item
$G$ is a normal distribution with mean $\xi$ and variance $\sigma^2$.
\end{bulllist}
\end{assumption}

\bigskip
Following the same procedure as above for $n\in\mathbb{N}$ yields
\begin{equation}\label{eq.k1}
\frac{Y_1+\cdots+Y_N}{n} \,\Big|\, N=n \sim {\rm N}\left( \mu_n, \sigma_n^2\right)
\end{equation}
 where
\[
\mu_n= \xi + \sigma\rho_1\Phi^{-1}(F_\alpha(n))
\quad\hbox{and}\quad
\sigma_n^2=\frac{\left(1-n\rho_1^2 \right)}{n}+ \frac{2}{n^2}\frac{\rho_2^2}{1-\rho_2^2}\left( \rho_2^{n-1}-1 \right).
\]
Again, the distribution of $S$ can be expressed in closed form which in turn allows a closed form expression for subsequent statistics of $S$.


\section*{Acknowledgements}

 Rosy Oh was supported by the Basic Science Research Program through the National Research Foundation of Korea (NRF) funded by the Ministry of Education (2019R1A6A1A11051177 and 2020R1I1A1A01067376).
Jae Youn Ahn was supported by an  NRF grant funded by the Korean Government (2020R1F1A1A01061202).

\bibliographystyle{apalike}


\appendix
\section{Appendix}

\begin{lemma}{}\label{lemma.1}
Consider $\alpha\in(0,1)$,
a random vector $(N, Y_1, \dots, Y_N)$ from a transformed copula-based CRM (Model \ref{mod.1})
satisfying Assumption \ref{assum.1}. Then, we have following properties:
\begin{itemize}
\item[i.]
For $n \in \mathbb{N}$, consider a random vector $\left(Z_0, Z_1, \cdots, Z_n \right)$
following a multivariate normal distribution with mean
$\big(0, \xi, \cdots, \xi \big)^\mathrm{T}$
and covariance matrix
$$
  {\rm diag}\left(1, \sigma, \dots, \sigma \right) \boldsymbol{\Sigma}_{\rho_1,\rho_2}^{[n,1]} {\rm diag}\left(1, \sigma, \dots, \sigma \right)
$$
where $\boldsymbol{J}_{n\times n}$ denotes an $n \times n$ matrix with entries $1$.
Then, the conditional distribution of $\left(Y_1, \cdots, Y_n \right)$ conditional on $N=n$
equals the conditional distribution of $\left(Z_1, \cdots, Z_n \right)$ conditional on
$Z_0=\Phi_{0,1}^{-1}\left( F_\alpha(n)\right)$ which satisfies
\begin{equation}
\begin{aligned}
  &\left(Z_1, \cdots, Z_n \right) \,|\, Z_0=\Phi_{0,1}^{-1}\left( F_\alpha(n)\right)\\
  &\quad\quad\quad\quad\quad\quad\quad\quad\sim{\rm MVN}\left(\left(\xi+\sigma \rho_1 \Phi_{0,1}^{-1}\left( F_\alpha(n)\right) \right)\boldsymbol{1}_n, \sigma^2\left( \boldsymbol{\Sigma}_{\rho_2}^{[n,1]} -\rho_1^2\boldsymbol{J}_{n\times n}\right) \right)
\end{aligned}
\end{equation}

\item[ii.]
Consider a random vector
$\left(X, Y_1, \cdots, Y_X\right)$ from a two part CRM (Model \ref{mod.a1}) and assume that $X\eqd N$ shares the same distribution function $F$ with $N\sim F$.
If we further assume $\rho_1=\rho_2^2$ and that $\mu_n$ and $\sigma_0^2$ in \eqref{eq.ahn2a} satisfying
$$
  \mu_n=\sigma\rho_1\Phi^{-1}\left( F_\alpha(n)\right)  \qquad \hbox{and} \qquad
  \sigma_0^2=\sigma^2\left(1-\rho_1^2\right)
$$
for $X=n\in\mathbb{N}$,
then the two random vectors $\left(N, Y_1, \cdots, Y_N \right)$ and $\left(X, Y_1, \cdots, Y_X\right)$ have the same distribution.
\end{itemize}
\end{lemma}

\bigskip
\begin{proof}
We first prove part i.
By \cite{oh2020copulaSAJ}, the matrix $\boldsymbol{\Sigma}_{\rho_1,\rho_2}^{[n,1]}$ is positive definite.
Furthermore, since the corresponding marginals and copula function of a multivariate normal distribution are normal distributions and Gaussian copula, respectively, the conditional density function of $\left(Z_1, \dots, Z_n \right)$
at point $\left(z_1, \dots, z_n \right)$ conditional on $Z_0=z_0$ equals
\begin{equation}\label{a1}
  c_{(n)}\left( \Phi_{0,1}(z_0), \Phi_{\xi, \sigma^2}\left( z_1 \right), \dots, \Phi_{\xi, \sigma^2}\left( z_n \right); \boldsymbol{\Sigma}_{\rho_1,\rho_2}^{[n,1]}\right)\prod\limits_{i=1}^{n}\phi_{\xi, \sigma^2}\left( z_i \right).
\end{equation}
On the other hand,
the conditional distribution of $\left(Z_1, \cdots, Z_n \right)$ conditional on $Z_0=z_0$ satisfies
  \begin{equation}\label{a2}
  \left(Z_1, \cdots, Z_n \right) \,|\, Z_0=z_0\sim{\rm MVN}\left(\left(\xi+\rho_1\sigma z_0 \right)\boldsymbol{1}_n, \sigma^2\left( \boldsymbol{\Sigma}_{\rho_2}^{[n,1]} -\rho_1^2\boldsymbol{J}_{n\times n}\right) \right)
  \end{equation}
Since \eqref{a1} and \eqref{a2} describe the same distribution,
the conditional distribution of $(Y_1, \dots, Y_n)$ conditional on $N=n$
and the conditional distribution of $(Z_1, \dots, Z_n)$ conditional on $Z_0=\Phi_{0,1}^{-1}\left( F_\alpha(n)\right)$ coincide which follows from \eqref{eq.ahn62}.
This proves part i.
The proof of part ii is immediate from part i.
\end{proof}

\newpage



\begin{table}
\caption{KL divergence $D(P,Q)$ from $P$ to $Q$ under various parameter settings for Gaussian copula  }\label{tb.kld.1}
\centering
\begin{tabular}{|c|c||c|c|c|c|c|}
\multicolumn{5}{l}{}\\
 \multicolumn{7}{l}{(a) $\alpha=0.25$ }\\
  \hline
   &  & \multicolumn{5}{c|}{$\zeta$} \\
  \hline
   &  & $0.1$ & $0.5$ & $1.0$ & $5.0$ & $10.0$ \\
  \hline\hline
   \multirow{7}{*}{$\theta$}
&	-0.951	&	4.532	&	1.932	&	0.998	&	0.149	&	0.072	\\ \cline{2-7}
&	-0.454	&	0.124	&	0.053	&	0.028	&	0.004	&	0.002	\\ \cline{2-7}
&	-0.156	&	0.012	&	0.005	&	0.003	&	0	&	0	\\ \cline{2-7}
&	0	&	0	&	0	&	0	&	0	&	0	\\ \cline{2-7}
&	0.156	&	0.012	&	0.005	&	0.003	&	0	&	0	\\ \cline{2-7}
&	0.454	&	0.124	&	0.054	&	0.028	&	0.004	&	0.002	\\ \cline{2-7}
&	0.951	&	4.526	&	1.932	&	1.004	&	0.149	&	0.071	\\ \cline{2-7}
  \hline
\multicolumn{5}{l}{}\\
\multicolumn{7}{l}{(b) $\alpha=0.5$  }\\
  \hline  
   &  & \multicolumn{5}{c|}{$\zeta$} \\
  \hline
   &  & $0.1$ & $0.5$ & $1.0$ & $5.0$ & $10.0$ \\
  \hline\hline
   \multirow{7}{*}{$\theta$}
&	-0.951	&	3.142	&	1.422	&	0.731	&	0.089	&	0.042	\\ \cline{2-7}
&	-0.454	&	0.087	&	0.039	&	0.020	&	0.002	&	0.001	\\ \cline{2-7}
&	-0.156	&	0.008	&	0.004	&	0.002	&	0	&	0	\\ \cline{2-7}
&	0	&	0	&	0	&	0	&	0	&	0	\\ \cline{2-7}
&	0.156	&	0.008	&	0.004	&	0.002	&	0	&	0	\\ \cline{2-7}
&	0.454	&	0.086	&	0.039	&	0.020	&	0.002	&	0.001	\\ \cline{2-7}
&	0.951	&	3.156	&	1.417	&	0.726	&	0.088	&	0.041	\\ \cline{2-7}  \hline
  \multicolumn{5}{l}{}\\
\multicolumn{7}{l}{(c) $\alpha=0.75$  }\\
  \hline  
   &  & \multicolumn{5}{c|}{$\zeta$} \\
  \hline
   &  & $0.1$ & $0.5$ & $1.0$ & $5.0$ & $10.0$ \\
  \hline\hline
   \multirow{7}{*}{$\theta$}
&	-0.951	&	4.962	&	2.246	&	1.153	&	0.153	&	0.072	\\ \cline{2-7}
&	-0.454	&	0.135	&	0.062	&	0.031	&	0.004	&	0.002	\\ \cline{2-7}
&	-0.156	&	0.013	&	0.006	&	0.003	&	0	&	0	\\ \cline{2-7}
&	0	&	0	&	0	&	0	&	0	&	0	\\ \cline{2-7}
&	0.156	&	0.013	&	0.006	&	0.003	&	0	&	0	\\ \cline{2-7}
&	0.454	&	0.136	&	0.062	&	0.032	&	0.004	&	0.002	\\ \cline{2-7}
&	0.951	&	4.955	&	2.244	&	1.150	&	0.153	&	0.072	\\ \cline{2-7}
  \hline
\multicolumn{5}{l}{}\\
\multicolumn{7}{l}{(d) $\alpha=1.0$  }\\
  \hline  
   &  & \multicolumn{5}{c|}{$\zeta$} \\
  \hline
   &  & $0.1$ & $0.5$ & $1.0$ & $5.0$ & $10.0$ \\
  \hline\hline
   \multirow{7}{*}{$\theta$}
&	-0.951	&	13.068	&	4.369	&	2.174	&	0.334	&	0.161	\\ \cline{2-7}
&	-0.454	&	0.359	&	0.119	&	0.060	&	0.009	&	0.004	\\ \cline{2-7}
&	-0.156	&	0.034	&	0.012	&	0.006	&	0.001	&	0	\\ \cline{2-7}
&	0	&	0	&	0	&	0	&	0	&	0	\\ \cline{2-7}
&	0.156	&	0.035	&	0.012	&	0.006	&	0.001	&	0	\\ \cline{2-7}
&	0.454	&	0.358	&	0.120	&	0.060	&	0.009	&	0.004	\\ \cline{2-7}
&	0.951	&	13.074	&	4.375	&	2.166	&	0.335	&	0.162	\\ \cline{2-7}
  \hline
\end{tabular}
\end{table}


\begin{table}
\caption{KL divergence $D(P,Q)$ from $P$ to $Q$ under various parameter settings for Student t copula  }\label{tb.kld.2}
\centering
\begin{tabular}{|c|c||c|c|c|c|c|}
\multicolumn{5}{l}{}\\
 \multicolumn{7}{l}{(a) $\alpha=0.25$ }\\
  \hline
   &  & \multicolumn{5}{c|}{$\zeta$} \\
  \hline
   &  & $0.1$ & $0.5$ & $1.0$ & $5.0$ & $10.0$ \\
  \hline\hline
   \multirow{7}{*}{$\theta$}
&	-0.951	&	2.559	&	1.450	&	0.878	&	0.165	&	0.081	\\ \cline{2-7}
&	-0.454	&	0.139	&	0.069	&	0.040	&	0.006	&	0.003	\\ \cline{2-7}
&	-0.156	&	0.026	&	0.017	&	0.012	&	0.002	&	0.001	\\ \cline{2-7}
&	0	&	0	&	0	&	0	&	0	&	0	\\ \cline{2-7}
&	0.156	&	0.025	&	0.017	&	0.012	&	0.002	&	0.001	\\ \cline{2-7}
&	0.454	&	0.138	&	0.069	&	0.040	&	0.006	&	0.003	\\ \cline{2-7}
&	0.951	&	2.562	&	1.447	&	0.879	&	0.166	&	0.081	\\ \cline{2-7}
  \hline
\multicolumn{5}{l}{}\\
\multicolumn{7}{l}{(b) $\alpha=0.5$  }\\
  \hline  
   &  & \multicolumn{5}{c|}{$\zeta$} \\
  \hline
   &  & $0.1$ & $0.5$ & $1.0$ & $5.0$ & $10.0$ \\
  \hline\hline
&	-0.951	&	2.025	&	1.079	&	0.638	&	0.103	&	0.047	\\ \cline{2-7}
&	-0.454	&	0.112	&	0.060	&	0.035	&	0.004	&	0.002	\\ \cline{2-7}
&	-0.156	&	0.026	&	0.018	&	0.011	&	0.001	&	0	\\ \cline{2-7}
&	0	&	0	&	0	&	0	&	0	&	0	\\ \cline{2-7}
&	0.156	&	0.026	&	0.017	&	0.012	&	0.001	&	0	\\ \cline{2-7}
&	0.454	&	0.113	&	0.060	&	0.035	&	0.004	&	0.002	\\ \cline{2-7}
&	0.951	&	2.027	&	1.078	&	0.642	&	0.103	&	0.048	\\ \cline{2-7}
\hline
  \multicolumn{5}{l}{}\\
\multicolumn{7}{l}{(c) $\alpha=0.75$  }\\
  \hline  
   &  & \multicolumn{5}{c|}{$\zeta$} \\
  \hline
   &  & $0.1$ & $0.5$ & $1.0$ & $5.0$ & $10.0$ \\
  \hline\hline
   \multirow{7}{*}{$\theta$}
&	-0.951	&	2.472	&	1.419	&	0.893	&	0.174	&	0.082	\\ \cline{2-7}
&	-0.454	&	0.151	&	0.086	&	0.051	&	0.007	&	0.003	\\ \cline{2-7}
&	-0.156	&	0.028	&	0.023	&	0.016	&	0.002	&	0.001	\\ \cline{2-7}
&	0	&	0	&	0	&	0	&	0	&	0	\\ \cline{2-7}
&	0.156	&	0.028	&	0.022	&	0.016	&	0.002	&	0.001	\\ \cline{2-7}
&	0.454	&	0.151	&	0.085	&	0.050	&	0.006	&	0.003	\\ \cline{2-7}
&	0.951	&	2.477	&	1.423	&	0.892	&	0.171	&	0.082	\\ \cline{2-7}
  \hline
\multicolumn{5}{l}{}\\
\multicolumn{7}{l}{(d) $\alpha=1.0$  }\\
  \hline  
   &  & \multicolumn{5}{c|}{$\zeta$} \\
  \hline
   &  & $0.1$ & $0.5$ & $1.0$ & $5.0$ & $10.0$ \\
  \hline\hline
   \multirow{7}{*}{$\theta$}
&	-0.951	&	4.576	&	2.454	&	1.568	&	0.361	&	0.181	\\ \cline{2-7}
&	-0.454	&	0.328	&	0.146	&	0.085	&	0.013	&	0.006	\\ \cline{2-7}
&	-0.156	&	0.057	&	0.033	&	0.024	&	0.004	&	0.002	\\ \cline{2-7}
&	0	&	0	&	0	&	0	&	0	&	0	\\ \cline{2-7}
&	0.156	&	0.057	&	0.032	&	0.023	&	0.004	&	0.002	\\ \cline{2-7}
&	0.454	&	0.327	&	0.146	&	0.085	&	0.013	&	0.006	\\ \cline{2-7}
&	0.951	&	4.572	&	2.447	&	1.565	&	0.360	&	0.181	\\ \cline{2-7}
  \hline
\end{tabular}
\end{table}


\begin{table}
\caption{KL divergence $D(P,Q)$ from $P$ to $Q$ under various parameter settings for Clayton copula}\label{tb.kld.3}
\centering
\begin{tabular}{|c|c||c|c|c|c|c|}
\multicolumn{5}{l}{}\\
 \multicolumn{7}{l}{(a) $\alpha=0.25$ }\\
  \hline
   &  & \multicolumn{5}{c|}{$\zeta$} \\
  \hline
   &  & $0.1$ & $0.5$ & $1.0$ & $5.0$ & $10.0$ \\
  \hline\hline
   \multirow{4}{*}{$\theta$}
&	0	&	0	&	0	&	0	&	0	&	0	\\ \cline{2-7}
&	0.222	&	0.019	&	0.013	&	0.008	&	0.001	&	0	\\ \cline{2-7}
&	0.857	&	0.192	&	0.131	&	0.083	&	0.010	&	0.004	\\ \cline{2-7}
&	8	&	5.041	&	3.459	&	2.306	&	0.462	&	0.224	\\ \cline{2-7}
  \hline
\multicolumn{5}{l}{}\\
\multicolumn{7}{l}{(b) $\alpha=0.5$  }\\
  \hline  
   &  & \multicolumn{5}{c|}{$\zeta$} \\
  \hline
   &  & $0.1$ & $0.5$ & $1.0$ & $5.0$ & $10.0$ \\
  \hline\hline
   \multirow{4}{*}{$\theta$}
&	0	&	0	&	0	&	0	&	0	&	0	\\ \cline{2-7}
&	0.222	&	0.020	&	0.013	&	0.008	&	0.001	&	0	\\ \cline{2-7}
&	0.857	&	0.196	&	0.131	&	0.082	&	0.007	&	0.003	\\ \cline{2-7}
&	8	&	3.953	&	2.697	&	1.770	&	0.313	&	0.143	\\ \cline{2-7}
  \hline
  \multicolumn{5}{l}{}\\
\multicolumn{7}{l}{(c) $\alpha=0.75$  }\\
  \hline  
   &  & \multicolumn{5}{c|}{$\zeta$} \\
  \hline
   &  & $0.1$ & $0.5$ & $1.0$ & $5.0$ & $10.0$ \\
  \hline\hline
   \multirow{4}{*}{$\theta$}
&	0	&	0	&	0	&	0	&	0	&	0	\\ \cline{2-7}
&	0.222	&	0.027	&	0.019	&	0.012	&	0.001	&	0	\\ \cline{2-7}
&	0.857	&	0.268	&	0.183	&	0.114	&	0.012	&	0.005	\\ \cline{2-7}
&	8	&	4.892	&	3.354	&	2.257	&	0.481	&	0.235	\\ \cline{2-7}
  \hline
\multicolumn{5}{l}{}\\
\multicolumn{7}{l}{(d) $\alpha=1.0$  }\\
  \hline  
   &  & \multicolumn{5}{c|}{$\zeta$} \\
  \hline
   &  & $0.1$ & $0.5$ & $1.0$ & $5.0$ & $10.0$ \\
  \hline\hline
   \multirow{4}{*}{$\theta$}
&	0	&	0	&	0	&	0	&	0	&	0	\\ \cline{2-7}
&	0.222	&	0.036	&	0.025	&	0.016	&	0.002	&	0.001	\\ \cline{2-7}
&	0.857	&	0.357	&	0.242	&	0.156	&	0.020	&	0.009	\\ \cline{2-7}
&	8	&	6.447	&	4.482	&	3.114	&	0.821	&	0.441	\\ \cline{2-7}
  \hline
\end{tabular}
\end{table}


\begin{table}
\caption{KL divergence $D(P,Q)$ from $P$ to $Q$ under various parameter settings for Gumbel copula}\label{tb.kld.4}
\centering
\begin{tabular}{|c|c||c|c|c|c|c|}
\multicolumn{5}{l}{}\\
 \multicolumn{7}{l}{(a) $\alpha=0.25$ }\\
  \hline
   &  & \multicolumn{5}{c|}{$\zeta$} \\
  \hline
   &  & $0.1$ & $0.5$ & $1.0$ & $5.0$ & $10.0$ \\
  \hline\hline
   \multirow{4}{*}{$\theta$}
&	1	&	0.247	&	0.168	&	0.107	&	0.013	&	0.006	\\ \cline{2-7}
&	1.111	&	0.293	&	0.197	&	0.126	&	0.016	&	0.007	\\ \cline{2-7}
&	1.429	&	0.434	&	0.295	&	0.188	&	0.024	&	0.010	\\ \cline{2-7}
&	5	&	2.749	&	1.872	&	1.226	&	0.211	&	0.097	\\ \cline{2-7}
  \hline
\multicolumn{5}{l}{}\\
\multicolumn{7}{l}{(b) $\alpha=0.5$  }\\
  \hline  
   &  & \multicolumn{5}{c|}{$\zeta$} \\
  \hline
   &  & $0.1$ & $0.5$ & $1.0$ & $5.0$ & $10.0$ \\
  \hline\hline
   \multirow{4}{*}{$\theta$}
&	1	&	0.247	&	0.166	&	0.103	&	0.009	&	0.004	\\ \cline{2-7}
&	1.111	&	0.289	&	0.196	&	0.123	&	0.011	&	0.004	\\ \cline{2-7}
&	1.429	&	0.422	&	0.282	&	0.179	&	0.017	&	0.006	\\ \cline{2-7}
&	5	&	2.239	&	1.513	&	0.977	&	0.144	&	0.062	\\ \cline{2-7}
  \hline
  \multicolumn{5}{l}{}\\
\multicolumn{7}{l}{(c) $\alpha=0.75$  }\\
  \hline  
   &  & \multicolumn{5}{c|}{$\zeta$} \\
  \hline
   &  & $0.1$ & $0.5$ & $1.0$ & $5.0$ & $10.0$ \\
  \hline\hline
   \multirow{4}{*}{$\theta$}
&	1	&	0.341	&	0.230	&	0.146	&	0.016	&	0.006	\\ \cline{2-7}
&	1.111	&	0.399	&	0.271	&	0.171	&	0.018	&	0.007	\\ \cline{2-7}
&	1.429	&	0.571	&	0.387	&	0.246	&	0.029	&	0.011	\\ \cline{2-7}
&	5	&	2.864	&	1.955	&	1.284	&	0.231	&	0.105	\\ \cline{2-7}
  \hline
\multicolumn{5}{l}{}\\
\multicolumn{7}{l}{(d) $\alpha=1.0$  }\\
  \hline  
   &  & \multicolumn{5}{c|}{$\zeta$} \\
  \hline
   &  & $0.1$ & $0.5$ & $1.0$ & $5.0$ & $10.0$ \\
  \hline\hline
   \multirow{4}{*}{$\theta$}
&	1	&	0.452	&	0.308	&	0.199	&	0.027	&	0.012	\\ \cline{2-7}
&	1.111	&	0.530	&	0.357	&	0.231	&	0.032	&	0.014	\\ \cline{2-7}
&	1.429	&	0.757	&	0.517	&	0.337	&	0.049	&	0.022	\\ \cline{2-7}
&	5	&	3.774	&	2.597	&	1.762	&	0.399	&	0.198	\\ \cline{2-7}
  \hline
\end{tabular}
\end{table}


\begin{table}
\caption{Spearman's rho values $\rho(P) \,/\, \rho(Q)$ under various parameter settings for Gaussian copula}\label{tb.rho.1}
\centering
\resizebox{\linewidth}{!}{
\begin{tabular}{|c|c||c|c|c|c|c|}
\multicolumn{5}{l}{}\\
 \multicolumn{7}{l}{(a) $\alpha=0.25$ }\\
  \hline
   &  & \multicolumn{5}{c|}{$\zeta$} \\
  \hline
   &  & $0.1$ & $0.5$ & $1.0$ & $5.0$ & $10.0$ \\
  \hline\hline
   \multirow{7}{*}{$\theta$}
&	-0.951	&	-0.946	/	-0.259	&	-0.946	/	-0.746	&	-0.946	/	-0.884	&	-0.946	/	-0.939	&	-0.946	/	-0.943	\\ \cline{2-7}
&	-0.454	&	-0.438	/	-0.148	&	-0.437	/	-0.358	&	-0.437	/	-0.411	&	-0.439	/	-0.438	&	-0.439	/	-0.437	\\ \cline{2-7}
&	-0.156	&	-0.150	/	-0.053	&	-0.149	/	-0.123	&	-0.150/	-0.147	&	-0.149	/	-0.149	&	-0.151	/	-0.149	\\ \cline{2-7}
&	0	&	0	/	-0.001	&	-0.001	/	0	&	-0.001	/	0	&	0	/	0.001	&	0.001	/	0.006	\\ \cline{2-7}
&	0.156	&	0.150	/	0.053	&	0.148	/	0.118	&	0.147	/	0.143	&	0.149	/	0.145	&	0.149	/	0.148	\\ \cline{2-7}
&	0.454	&	0.436	/	0.156	&	0.438	/	0.366	&	0.437	/	0.413	&	0.438	/	0.437	&	0.437	/	0.437	\\ \cline{2-7}
&	0.951	&	0.946	/	0.258	&	0.947	/	0.747	&	0.947	/	0.881	&	0.946	/	0.934	&	0.946	/	0.945	\\ \cline{2-7}
  \hline
\multicolumn{5}{l}{}\\
\multicolumn{7}{l}{(b) $\alpha=0.5$  }\\
  \hline  
   &  & \multicolumn{5}{c|}{$\zeta$} \\
  \hline
   &  & $0.1$ & $0.5$ & $1.0$ & $5.0$ & $10.0$ \\
  \hline\hline
   \multirow{7}{*}{$\theta$}
&	-0.951	&	-0.946	/	-0.256	&	-0.946	/	-0.745	&	-0.946	/	-0.876	&	-0.946	/	-0.934	&	-0.947	/	-0.942	\\ \cline{2-7}
&	-0.454	&	-0.437	/	-0.128	&	-0.438	/	-0.304	&	-0.439	/	-0.369	&	-0.437	/	-0.428	&	-0.438	/	-0.433	\\ \cline{2-7}
&	-0.156	&	-0.150	/	-0.041	&	-0.150	/	-0.103	&	-0.148	/	-0.121	&	-0.149	/	-0.149	&	-0.149	/	-0.149	\\ \cline{2-7}
&	0	&	0	/	0	&	0.001	/	0	&	0.001	/	-0.001	&	0	/	-0.003	&	0	/	-0.004	\\ \cline{2-7}
&	0.156	&	0.149	/	0.038	&	0.149	/	0.104	&	0.149	/	0.121	&	0.148	/	0.145	&	0.149	/	0.147	\\ \cline{2-7}
&	0.454	&	0.439	/	0.125	&	0.437	/	0.308	&	0.438	/	0.364	&	0.436	/	0.426	&	0.437	/	0.435	\\ \cline{2-7}
&	0.951	&	0.946	/	0.256	&	0.946	/	0.749	&	0.946	/	0.881	&	0.946	/	0.938	&	0.946	/	0.944	\\ \cline{2-7}
  \hline
  \multicolumn{5}{l}{}\\
\multicolumn{7}{l}{(c) $\alpha=0.75$  }\\
  \hline  
   &  & \multicolumn{5}{c|}{$\zeta$} \\
  \hline
   &  & $0.1$ & $0.5$ & $1.0$ & $5.0$ & $10.0$ \\
  \hline\hline
   \multirow{7}{*}{$\theta$}
&	-0.951	&	-0.946	/	-0.26	&	-0.946	/	-0.739	&	-0.946	/	-0.872	&	-0.947	/	-0.936	&	-0.946	/	-0.939	\\ \cline{2-7}
&	-0.454	&	-0.437	/	-0.104	&	-0.437	/	-0.280	&	-0.436	/	-0.353	&	-0.436	/	-0.425	&	-0.438	/	-0.430\\ \cline{2-7}
&	-0.156	&	-0.150	/	-0.035	&	-0.148	/	-0.097	&	-0.150	/	-0.116	&-0.150/	-0.144	&	-0.149	/	-0.149	\\ \cline{2-7}
&	0	&	0	/	0.004	&	-0.002	/	0.003	&	-0.001	/	0.001	&	0.002	/	-0.001	&	0.001	/	0.003	\\ \cline{2-7}
&	0.156	&	0.148	/	0.034	&	0.150	/	0.087	&	0.149	/	0.116	&	0.148	/	0.144	&	0.150	/	0.137	\\ \cline{2-7}
&	0.454	&	0.437	/	0.105	&	0.438	/	0.278	&	0.436	/	0.356	&	0.437	/	0.420	&	0.437	/	0.434	\\ \cline{2-7}
&	0.951	&	0.947	/	0.263	&	0.946	/	0.737	&	0.946	/	0.871	&	0.946	/	0.933	&	0.947	/	0.943	\\ \cline{2-7}
  \hline
\multicolumn{5}{l}{}\\
\multicolumn{7}{l}{(d) $\alpha=1.0$  }\\
  \hline  
   &  & \multicolumn{5}{c|}{$\zeta$} \\
  \hline
   &  & $0.1$ & $0.5$ & $1.0$ & $5.0$ & $10.0$ \\
  \hline\hline
   \multirow{7}{*}{$\theta$}
 &	-0.951	&	-0.946	/	-0.259	&	-0.946	/	-0.737	&	-0.947	/	-0.873	&	-0.946	/	-0.934	&	-0.946	/	-0.942	\\ \cline{2-7}
&	-0.454	&	-0.439	/	-0.099	&	-0.438	/	-0.271	&	-0.438	/	-0.355	&	-0.438	/	-0.425	&	-0.438	/	-0.430	\\ \cline{2-7}
&	-0.156	&	-0.152	/	-0.027	&	-0.149	/	-0.089	&	-0.151	/	-0.114	&	-0.150	/	-0.141	&	-0.148	/	-0.144	\\ \cline{2-7}
&	0	&	0.001	/	0.002	&	-0.001	/	0.003	&	-0.001	/	-0.006	&	-0.001	/	0	&	0	/	-0.002	\\ \cline{2-7}
&	0.156	&	0.149	/	0.033	&	0.150	/	0.092	&	0.149	/	0.122	&	0.150	/	0.144	&	0.150	/	0.148	\\ \cline{2-7}
&	0.454	&	0.437	/	0.099	&	0.437	/	0.276	&	0.437	/	0.348	&	0.437	/	0.423	&	0.438	/	0.431	\\ \cline{2-7}
&	0.951	&	0.947	/	0.249	&	0.946	/	0.737	&	0.946	/	0.877	&	0.946	/	0.937	&	0.947	/	0.941	\\ \cline{2-7}
   \hline
\end{tabular}
} 
\end{table}


\begin{table}
\caption{Spearman's rho values $\rho(P) \,/\, \rho(Q)$ under various parameter settings for Student t copula}\label{tb.rho.2}
\centering
\resizebox{\linewidth}{!}{
\begin{tabular}{|c|c||c|c|c|c|c|}
\multicolumn{5}{l}{}\\
 \multicolumn{7}{l}{(a) $\alpha=0.25$ }\\
  \hline
   &  & \multicolumn{5}{c|}{$\zeta$} \\
  \hline
   &  & $0.1$ & $0.5$ & $1.0$ & $5.0$ & $10.0$ \\
  \hline\hline
   \multirow{7}{*}{$\theta$}
&	-0.951	&	-0.942	/	-0.258	&	-0.942	/	-0.743	&	-0.942	/	-0.879	&	-0.942	/	-0.935	&	-0.942	/	-0.938	\\ \cline{2-7}
&	-0.454	&	-0.427	/	-0.144	&	-0.427	/	-0.353	&	-0.427	/	-0.407	&	-0.427	/	-0.425	&	-0.426	/	-0.428	\\ \cline{2-7}
&	-0.156	&	-0.144	/	-0.050	&	-0.145	/	-0.115	&	-0.144	/	-0.136	&	-0.144	/	-0.143	&	-0.148	/	-0.143	\\ \cline{2-7}
&	0	&	-0.001	/	0.001	&	-0.002	/	-0.002	&	-0.002	/	-0.005	&	0	/	0.004	&	0	/	-0.004	\\ \cline{2-7}
&	0.156	&	0.147	/	0.053	&	0.146	/	0.121	&	0.146	/	0.141	&	0.147	/	0.144	&	0.145	/	0.146	\\ \cline{2-7}
&	0.454	&	0.428	/	0.145	&	0.428	/	0.352	&	0.426	/	0.407	&	0.429	/	0.429	&	0.427	/	0.430	\\ \cline{2-7}
&	0.951	&	0.942	/	0.256	&	0.942	/	0.742	&	0.942	/	0.871	&	0.942	/	0.936	&	0.943	/	0.936	\\ \cline{2-7}
 \hline
\multicolumn{5}{l}{}\\
\multicolumn{7}{l}{(b) $\alpha=0.5$  }\\
  \hline  
   &  & \multicolumn{5}{c|}{$\zeta$} \\
  \hline
   &  & $0.1$ & $0.5$ & $1.0$ & $5.0$ & $10.0$ \\
  \hline\hline
   \multirow{7}{*}{$\theta$}
&	-0.951	&	-0.942	/	-0.258	&	-0.942	/	-0.739	&	-0.942	/	-0.869	&	-0.942	/	-0.930	&	-0.942	/	-0.937	\\ \cline{2-7}
&	-0.454	&	-0.426	/	-0.122	&	-0.427	/	-0.302	&	-0.426	/	-0.364	&	-0.427	/	-0.419	&	-0.425	/	-0.425	\\ \cline{2-7}
&	-0.156	&	-0.146	/	-0.043	&	-0.146	/	-0.106	&	-0.146	/	-0.120	&	-0.145	/	-0.144	&	-0.145	/	-0.141	\\ \cline{2-7}
&	0	&	0	/	-0.002	&	0.001	/	-0.003	&	-0.001	/	0	&	-0.001	/	-0.004	&	0.001	/	0.002	\\ \cline{2-7}
&	0.156	&	0.145	/	0.048	&	0.145	/	0.101	&	0.147	/	0.123	&	0.146	/	0.141	&	0.145	/	0.141	\\ \cline{2-7}
&	0.454	&	0.426	/	0.125	&	0.426	/	0.304	&	0.427	/	0.366	&	0.427	/	0.418	&	0.427	/	0.420	\\ \cline{2-7}
&	0.951	&	0.942	/	0.249	&	0.942	/	0.737	&	0.942	/	0.871	&	0.942	/	0.935	&	0.942	/	0.933	\\ \cline{2-7}
  \hline
  \multicolumn{5}{l}{}\\
\multicolumn{7}{l}{(c) $\alpha=0.75$  }\\
  \hline  
   &  & \multicolumn{5}{c|}{$\zeta$} \\
  \hline
   &  & $0.1$ & $0.5$ & $1.0$ & $5.0$ & $10.0$ \\
  \hline\hline
   \multirow{7}{*}{$\theta$}
&	-0.951	&	-0.942	/	-0.256	&	-0.942	/	-0.731	&	-0.942	/	-0.865	&	-0.942	/	-0.932	&	-0.942	/	-0.935	\\ \cline{2-7}
&	-0.454	&	-0.428	/	-0.117	&	-0.428	/	-0.286	&	-0.426	/	-0.351	&	-0.428	/	-0.415	&	-0.428	/	-0.421	\\ \cline{2-7}
&	-0.156	&	-0.145	/	-0.039	&	-0.147	/	-0.096	&	-0.145	/	-0.121	&	-0.145	/	-0.144	&	-0.145	/	-0.145	\\ \cline{2-7}
&	0	&	0	/	0.001	&	0	/	0	&	0.001	/	0	&	0	/	0.004	&	0.001	/	0.003	\\ \cline{2-7}
&	0.156	&	0.146	/	0.035	&	0.146	/	0.089	&	0.146	/	0.121	&	0.146	/	0.140	&	0.144	/	0.146	\\ \cline{2-7}
&	0.454	&	0.429	/	0.111	&	0.429	/	0.277	&	0.428	/	0.352	&	0.429	/	0.411	&	0.427	/	0.424	\\ \cline{2-7}
&	0.951	&	0.942	/	0.257	&	0.942	/	0.725	&	0.942	/	0.866	&	0.942	/	0.932	&	0.942	/	0.934	\\ \cline{2-7}
  \hline
\multicolumn{5}{l}{}\\
\multicolumn{7}{l}{(d) $\alpha=1.0$  }\\
  \hline  
   &  & \multicolumn{5}{c|}{$\zeta$} \\
  \hline
   &  & $0.1$ & $0.5$ & $1.0$ & $5.0$ & $10.0$ \\
  \hline\hline
   \multirow{7}{*}{$\theta$}
&	-0.951	&	-0.942	/	-0.255	&	-0.942	/	-0.730	&	-0.942	/	-0.867	&	-0.942	/	-0.930	&	-0.942	/	-0.935	\\ \cline{2-7}
&	-0.454	&	-0.428	/	-0.108	&	-0.428	/	-0.283	&	-0.428	/	-0.355	&	-0.427	/	-0.415	&	-0.428	/	-0.417	\\ \cline{2-7}
&	-0.156	&	-0.147	/	-0.039	&	-0.146	/	-0.095	&	-0.146	/	-0.118	&	-0.146	/	-0.141	&	-0.145	/	-0.136	\\ \cline{2-7}
&	0	&	0	/	0.001	&	0.001	/	0.001	&	-0.001	/	0.001	&	0	/	0.003	&	-0.002	/	-0.001	\\ \cline{2-7}
&	0.156	&	0.146	/	0.033	&	0.147	/	0.096	&	0.147	/	0.119	&	0.147	/	0.143	&	0.146	/	0.143	\\ \cline{2-7}
&	0.454	&	0.428	/	0.115	&	0.428	/	0.282	&	0.427	/	0.354	&	0.426	/	0.416	&	0.427	/	0.421	\\ \cline{2-7}
&	0.951	&	0.942	/	0.252	&	0.942	/	0.729	&	0.942	/	0.866	&	0.942	/	0.933	&	0.942	/	0.939	\\ \cline{2-7}
   \hline
\end{tabular}
} 
\end{table}


\begin{table}
\caption{Spearman's rho values $\rho(P) \,/\, \rho(Q)$ under various parameter settings for Clayton copula}\label{tb.rho.3}
\centering
\resizebox{\linewidth}{!}{
\begin{tabular}{|c|c||c|c|c|c|c|}
\multicolumn{5}{l}{}\\
 \multicolumn{7}{l}{(a) $\alpha=0.25$ }\\
  \hline
   &  & \multicolumn{5}{c|}{$\zeta$} \\
  \hline
   &  & $0.1$ & $0.5$ & $1.0$ & $5.0$ & $10.0$ \\
  \hline\hline
   \multirow{4}{*}{$\theta$}
&	0	&	-0.001	/	0.004	&	0	/	0	&	-0.001	/	-0.001	&	0	/	0.001	&	-0.001	/	-0.004	\\ \cline{2-7}
&	0.222	&	0.148	/	0.035	&	0.149	/	0.117	&	0.149	/	0.153	&	0.149	/	0.158	&	0.149	/	0.156	\\ \cline{2-7}
&	0.857	&	0.434	/	0.117	&	0.435	/	0.375	&	0.435	/	0.454	&	0.434	/	0.461	&	0.434	/	0.454	\\ \cline{2-7}
&	8	&	0.941	/	0.262	&	0.941	/	0.738	&	0.941	/	0.874	&	0.941	/	0.945	&	0.941	/	0.939	\\ \cline{2-7}
  \hline
\multicolumn{5}{l}{}\\
\multicolumn{7}{l}{(b) $\alpha=0.5$  }\\
  \hline  
   &  & \multicolumn{5}{c|}{$\zeta$} \\
  \hline
   &  & $0.1$ & $0.5$ & $1.0$ & $5.0$ & $10.0$ \\
  \hline\hline
   \multirow{4}{*}{$\theta$}
&	0	&	-0.001	/	-0.006	&	0	/	0.001	&	0.001	/	0.001	&	0	/	-0.003	&	0	/	-0.001	\\ \cline{2-7}
&	0.222	&	0.149	/	0.021	&	0.148	/	0.071	&	0.149	/	0.104	&	0.149	/	0.147	&	0.150	/	0.151	\\ \cline{2-7}
&	0.857	&	0.434	/	0.060	&	0.435	/	0.241	&	0.433	/	0.337	&	0.433	/	0.427	&	0.435	/	0.429	\\ \cline{2-7}
&	8	&	0.941	/	0.248	&	0.941	/	0.729	&	0.941	/	0.864	&	0.941	/	0.928	&	0.941	/	0.936	\\ \cline{2-7}
  \hline
  \multicolumn{5}{l}{}\\
\multicolumn{7}{l}{(c) $\alpha=0.75$  }\\
  \hline  
   &  & \multicolumn{5}{c|}{$\zeta$} \\
  \hline
   &  & $0.1$ & $0.5$ & $1.0$ & $5.0$ & $10.0$ \\
  \hline\hline
   \multirow{4}{*}{$\theta$}
&	0	&	-0.002	/	0	&	0	/	-0.001	&	-0.001	/	-0.005	&	0.002	/	0.003	&	0	/	-0.003	\\ \cline{2-7}
&	0.222	&	0.148	/	0.013	&	0.147	/	0.049	&	0.150	/	0.080	&	0.148	/	0.136	&	0.150	/	0.136	\\ \cline{2-7}
&	0.857	&	0.434	/	0.030	&	0.435	/	0.162	&	0.435	/	0.271	&	0.434	/	0.393	&	0.433	/	0.411	\\ \cline{2-7}
&	8	&	0.941	/	0.202	&	0.941	/	0.698	&	0.941	/	0.839	&	0.941	/	0.920	&	0.941	/	0.924	\\ \cline{2-7}
  \hline
\multicolumn{5}{l}{}\\
\multicolumn{7}{l}{(d) $\alpha=1.0$  }\\
  \hline  
   &  & \multicolumn{5}{c|}{$\zeta$} \\
  \hline
   &  & $0.1$ & $0.5$ & $1.0$ & $5.0$ & $10.0$ \\
  \hline\hline
   \multirow{4}{*}{$\theta$}
 &	0	&	0.001	/	0.003	&	0	/	0.004	&	0.001	/	-0.004	&	-0.001	/	0.009	&	-0.001	/	-0.001	\\ \cline{2-7}
&	0.222	&	0.149	/	0.003	&	0.150	/	0.031	&	0.148	/	0.062	&	0.149	/	0.123	&	0.149	/	0.134	\\ \cline{2-7}
&	0.857	&	0.435	/	0.012	&	0.434	/	0.111	&	0.435	/	0.214	&	0.435	/	0.368	&	0.433	/	0.390	\\ \cline{2-7}
&	8	&	0.941	/	0.067	&	0.941	/	0.611	&	0.941	/	0.798	&	0.941	/	0.904	&	0.941	/	0.924	\\ \cline{2-7}
    \hline
\end{tabular}
} 
\end{table}


\begin{table}
\caption{Spearman's rho values $\rho(P) \,/\, \rho(Q)$ under various parameter settings for Gumbel copula}\label{tb.rho.4}
\centering
\resizebox{\linewidth}{!}{
\begin{tabular}{|c|c||c|c|c|c|c|}
\multicolumn{5}{l}{}\\
 \multicolumn{7}{l}{(a) $\alpha=0.25$ }\\
  \hline
   &  & \multicolumn{5}{c|}{$\zeta$} \\
  \hline
   &  & $0.1$ & $0.5$ & $1.0$ & $5.0$ & $10.0$ \\
  \hline\hline
   \multirow{4}{*}{$\theta$}
&	1	&	0.479	/	0.137	&	0.480	/	0.414	&	0.478	/	0.508	&	0.479	/	0.506	&	0.479	/	0.499	\\ \cline{2-7}
&	1.111	&	0.510	/	0.146	&	0.510	/	0.439	&	0.509	/	0.534	&	0.510	/	0.535	&	0.508	/	0.533	\\ \cline{2-7}
&	1.429	&	0.584	/	0.173	&	0.583	/	0.516	&	0.586	/	0.613	&	0.585	/	0.602	&	0.585	/	0.602	\\ \cline{2-7}
&	5	&	0.884	/	0.252	&	0.885	/	0.734	&	0.885	/	0.847	&	0.885	/	0.891	&	0.884	/	0.896	\\ \cline{2-7}
  \hline
\multicolumn{5}{l}{}\\
\multicolumn{7}{l}{(b) $\alpha=0.5$  }\\
  \hline  
   &  & \multicolumn{5}{c|}{$\zeta$} \\
  \hline
   &  & $0.1$ & $0.5$ & $1.0$ & $5.0$ & $10.0$ \\
  \hline\hline
   \multirow{4}{*}{$\theta$}
&	1	&	0.479	/	0.072	&	0.477	/	0.270	&	0.478	/	0.374	&	0.477	/	0.469	&	0.479	/	0.472	\\ \cline{2-7}
&	1.111	&	0.510	/	0.083	&	0.511	/	0.296	&	0.508	/	0.414	&	0.509	/	0.497	&	0.511	/	0.504	\\ \cline{2-7}
&	1.429	&	0.585	/	0.096	&	0.583	/	0.360	&	0.586	/	0.491	&	0.585	/	0.572	&	0.585	/	0.585	\\ \cline{2-7}
&	5	&	0.885	/	0.231	&	0.884	/	0.695	&	0.884	/	0.820	&	0.885	/	0.874	&	0.885	/	0.877	\\ \cline{2-7}
  \hline
  \multicolumn{5}{l}{}\\
\multicolumn{7}{l}{(c) $\alpha=0.75$  }\\
  \hline  
   &  & \multicolumn{5}{c|}{$\zeta$} \\
  \hline
   &  & $0.1$ & $0.5$ & $1.0$ & $5.0$ & $10.0$ \\
  \hline\hline
   \multirow{4}{*}{$\theta$}
&	1	&	0.478	/	0.041	&	0.479	/	0.185	&	0.478	/	0.301	&	0.480	/	0.441	&	0.478	/	0.455	\\ \cline{2-7}
&	1.111	&	0.509	/	0.043	&	0.510	/	0.197	&	0.509	/	0.333	&	0.509	/	0.464	&	0.510	/	0.477	\\ \cline{2-7}
&	1.429	&	0.585	/	0.048	&	0.584	/	0.259	&	0.584	/	0.401	&	0.585	/	0.542	&	0.584	/	0.558	\\ \cline{2-7}
&	5	&	0.885	/	0.151	&	0.885	/	0.603	&	0.884	/	0.772	&	0.885	/	0.857	&	0.885	/	0.870	\\ \cline{2-7}
  \hline
\multicolumn{5}{l}{}\\
\multicolumn{7}{l}{(d) $\alpha=1.0$  }\\
  \hline  
   &  & \multicolumn{5}{c|}{$\zeta$} \\
  \hline
   &  & $0.1$ & $0.5$ & $1.0$ & $5.0$ & $10.0$ \\
  \hline\hline
   \multirow{4}{*}{$\theta$}
&	1	&	0.479	/	0.013	&	0.479	/	0.122	&	0.478	/	0.247	&	0.476	/	0.411	&	0.479	/	0.430	\\ \cline{2-7}
&	1.111	&	0.509	/	0.014	&	0.510	/	0.130	&	0.508	/	0.265	&	0.509	/	0.444	&	0.510	/	0.465	\\ \cline{2-7}
&	1.429	&	0.583	/	0.016	&	0.585	/	0.168	&	0.584	/	0.328	&	0.584	/	0.512	&	0.585	/	0.538	\\ \cline{2-7}
&	5	&	0.884	/	0.044	&	0.884	/	0.466	&	0.884	/	0.713	&	0.885	/	0.832	&	0.884	/	0.848	\\ \cline{2-7}
    \hline
\end{tabular}
} 
\end{table}


\begin{table}
\caption{KL divergence $D(P,Q)$ from $P$ to $Q$ under various parameter settings for Gaussian copula of dimension $3$}\label{tb.kld.d3.1}
\centering
\begin{tabular}{|c|c||c|c|c|c|c|}
\multicolumn{5}{l}{}\\
 \multicolumn{7}{l}{(a) $\alpha=0.25$ }\\
  \hline
   &  & \multicolumn{5}{c|}{$\zeta$} \\
  \hline
   &  & $0.1$ & $0.5$ & $1.0$ & $5.0$ & $10.0$ \\
  \hline\hline
   \multirow{5}{*}{$\theta$}
&	0	&	0	&	0	&	0	&	0	&	0	\\ \cline{2-7}
&	0.156	&	0.022	&	0.009	&	0.005	&	0.001	&	0	\\ \cline{2-7}
&	0.454	&	0.190	&	0.081	&	0.042	&	0.006	&	0.003	\\ \cline{2-7}
&	0.951	&	6.091	&	2.597	&	1.347	&	0.200	&	0.096	\\ \cline{2-7}
\hline
\multicolumn{5}{l}{}\\
\multicolumn{7}{l}{(b) $\alpha=0.5$  }\\
  \hline  
   &  & \multicolumn{5}{c|}{$\zeta$} \\
  \hline
   &  & $0.1$ & $0.5$ & $1.0$ & $5.0$ & $10.0$ \\
  \hline\hline
   \multirow{5}{*}{$\theta$}
&	0	&	0	&	0	&	0	&	0	&	0	\\ \cline{2-7}
&	0.156	&	0.015	&	0.007	&	0.003	&	0	&	0	\\ \cline{2-7}
&	0.454	&	0.132	&	0.059	&	0.030	&	0.004	&	0.002	\\ \cline{2-7}
&	0.951	&	4.224	&	1.907	&	0.977	&	0.118	&	0.056	\\ \cline{2-7}
\hline
  \multicolumn{5}{l}{}\\
\multicolumn{7}{l}{(c) $\alpha=0.75$  }\\
  \hline  
   &  & \multicolumn{5}{c|}{$\zeta$} \\
  \hline
   &  & $0.1$ & $0.5$ & $1.0$ & $5.0$ & $10.0$ \\
  \hline\hline
   \multirow{5}{*}{$\theta$}
&	0	&	0	&	0	&	0	&	0	&	0	\\ \cline{2-7}
&	0.156	&	0.023	&	0.011	&	0.005	&	0.001	&	0	\\ \cline{2-7}
&	0.454	&	0.207	&	0.094	&	0.048	&	0.006	&	0.003	\\ \cline{2-7}
&	0.951	&	6.675	&	3.023	&	1.553	&	0.204	&	0.097	\\ \cline{2-7}
  \hline
\multicolumn{5}{l}{}\\
\multicolumn{7}{l}{(d) $\alpha=1.0$  }\\
  \hline  
   &  & \multicolumn{5}{c|}{$\zeta$} \\
  \hline
   &  & $0.1$ & $0.5$ & $1.0$ & $5.0$ & $10.0$ \\
  \hline\hline
   \multirow{5}{*}{$\theta$}
&	0	&	0	&	0	&	0	&	0	&	0	\\ \cline{2-7}
&	0.156	&	0.061	&	0.020	&	0.010	&	0.002	&	0.001	\\ \cline{2-7}
&	0.454	&	0.547	&	0.182	&	0.092	&	0.014	&	0.007	\\ \cline{2-7}
&	0.951	&	17.564	&	5.877	&	2.914	&	0.451	&	0.218	\\ \cline{2-7}  \hline
\end{tabular}
\end{table}

\begin{table}
\caption{KL divergence $D(P,Q)$ from $P$ to $Q$ under various parameter settings for Clayton copula of dimension $3$}\label{tb.kld.d3.2}
\centering
\begin{tabular}{|c|c||c|c|c|c|c|}
\multicolumn{5}{l}{}\\
 \multicolumn{7}{l}{(a) $\alpha=0.25$ }\\
  \hline
   &  & \multicolumn{5}{c|}{$\zeta$} \\
  \hline
   &  & $0.1$ & $0.5$ & $1.0$ & $5.0$ & $10.0$ \\
  \hline\hline
   \multirow{4}{*}{$\theta$}
&	0	&	0	&	0	&	0	&	0	&	0	\\ \cline{2-7}
&	0.222	&	0.033	&	0.023	&	0.014	&	0.002	&	0.001	\\ \cline{2-7}
&	0.857	&	0.298	&	0.203	&	0.130	&	0.016	&	0.007	\\ \cline{2-7}
&	8	&	8.277	&	5.678	&	3.806	&	0.795	&	0.381	\\ \cline{2-7}
  \hline
\multicolumn{5}{l}{}\\
\multicolumn{7}{l}{(b) $\alpha=0.5$  }\\
  \hline  
   &  & \multicolumn{5}{c|}{$\zeta$} \\
  \hline
   &  & $0.1$ & $0.5$ & $1.0$ & $5.0$ & $10.0$ \\
  \hline\hline
   \multirow{4}{*}{$\theta$}
&	0	&	0	&	0	&	0	&	0	&	0	\\ \cline{2-7}
&	0.222	&	0.033	&	0.022	&	0.014	&	0.001	&	0	\\ \cline{2-7}
&	0.857	&	0.270	&	0.182	&	0.114	&	0.011	&	0.004	\\ \cline{2-7}
&	8	&	4.757	&	3.257	&	2.176	&	0.435	&	0.206	\\ \cline{2-7}
\hline
  \multicolumn{5}{l}{}\\
\multicolumn{7}{l}{(c) $\alpha=0.75$  }\\
  \hline  
   &  & \multicolumn{5}{c|}{$\zeta$} \\
  \hline
   &  & $0.1$ & $0.5$ & $1.0$ & $5.0$ & $10.0$ \\
  \hline\hline
   \multirow{4}{*}{$\theta$}
&	0	&	0	&	0	&	0	&	0	&	0	\\ \cline{2-7}
&	0.222	&	0.047	&	0.032	&	0.020	&	0.002	&	0.001	\\ \cline{2-7}
&	0.857	&	0.369	&	0.251	&	0.160	&	0.018	&	0.007	\\ \cline{2-7}
&	8	&	5.235	&	3.611	&	2.466	&	0.583	&	0.303	\\ \cline{2-7}
  \hline
\multicolumn{5}{l}{}\\
\multicolumn{7}{l}{(d) $\alpha=1.0$  }\\
  \hline  
   &  & \multicolumn{5}{c|}{$\zeta$} \\
  \hline
   &  & $0.1$ & $0.5$ & $1.0$ & $5.0$ & $10.0$ \\
  \hline\hline
   \multirow{4}{*}{$\theta$}
&	0	&	0	&	0	&	0	&	0	&	0	\\ \cline{2-7}
&	0.222	&	0.061	&	0.042	&	0.027	&	0.003	&	0.001	\\ \cline{2-7}
&	0.857	&	0.489	&	0.335	&	0.214	&	0.032	&	0.014	\\ \cline{2-7}
&	8	&	6.816	&	4.771	&	3.383	&	0.976	&	0.546	\\ \cline{2-7}
  \hline
\end{tabular}
\end{table}

\end{document}